\documentclass[preprint,review,12pt]{elsarticle}

\usepackage{graphicx,xcolor,tabularx}
\usepackage{epsfig}
\usepackage{amssymb}
\usepackage{pdfpages}
\usepackage{natbib}
\bibpunct{(}{)}{,}{a}{,}{,}
\newcommand{\bmath}[1]{\mbox{\boldmath$#1$}}
\usepackage{amsmath}
\usepackage{stmaryrd}
\usepackage{textcomp,tipa,booktabs,rotating}

\usepackage[english]{babel}
\usepackage[a4paper,centering=true,scale=0.75]{geometry}
\usepackage{amsthm, dsfont}
\usepackage{nicefrac}
\usepackage[bitstream-charter, cal=cmcal]{mathdesign}

\usepackage{bm}
\usepackage{subdepth}
\usepackage[utf8]{inputenc}
\usepackage{csquotes}
\usepackage{setspace}
\usepackage{enumerate}
\usepackage{relsize}
\usepackage{here}
\usepackage[bookmarks=true]{hyperref}

\definecolor{blue}{rgb}{0,0,1}
\definecolor{red}{rgb}{1,0,0}
\definecolor{green}{rgb}{0.2,0.6,0.2}

\newcommand{\blanco}[1]{}

\renewcommand{\boldsymbol}{\bm}

\renewcommand{\[}{\begin{equation*}}
\renewcommand{\]}{\end{equation*}}

\makeatletter
\def\ps@pprintTitle{%
 \let\@oddhead\@empty
 \let\@evenhead\@empty
 \def\@oddfoot{}%
 \let\@evenfoot\@oddfoot}
\makeatother

\journal{---------------}

\begin{document}

\begin{frontmatter}

\title{Classification of Functional Data with $k$-Nearest-Neighbor Ensembles by Fitting Constrained Multinomial Logit Models}
\tnotetext[]{Please see the arXiv comments for information about supplementary material.}
 \author[siemens,lmu]{Karen Fuchs\corref{cor1}\fnref{fnref1}}
 \ead{karenfuchs@gmx.de}
 \author[lmu]{Wolfgang P\"{o}{\ss}necker\fnref{fnref1}}
 \ead{Wolfgang.Poessnecker@stat.uni-muenchen.de}
 \author[lmu]{Gerhard Tutz}
 \ead{gerhard.tutz@stat.uni-muenchen.de}
 \cortext[cor1]{Corresponding author}
 \fntext[fnref1]{contributed equally}
 \address[siemens]{Siemens AG, CT RDA SII CPS-DE, Otto-Hahn-Ring 6, D-81739  Munich, Germany}
 \address[lmu]{Department of Statistics, Seminar of Applied Stochastics, Ludwig-Maximilians-Universit\"{a}t M\"{u}nchen, Akademiestr. 1, D-80799 Munich, Germany}

\title{Classification of Functional Data with $k$-Nearest-Neighbor Ensembles by Fitting Constrained Multinomial Logit Models}

\begin{abstract}
During the last decades, many methods for the analysis of functional data including classification methods have been developed. Nonetheless, there are issues that have not been adressed satisfactorily by currently available methods, as, for example, feature selection combined with variable selection when using multiple functional covariates. In this paper, a functional ensemble is combined with a penalized and constrained multinomial logit model. It is shown that this synthesis yields a powerful classification tool for functional data (possibly mixed with non-functional predictors), which also provides automatic variable selection. The choice of an appropriate, sparsity-inducing penalty allows to estimate most model coefficients to exactly zero, and permits class-specific coefficients in multi-class problems, such that feature selection is obtained. An additional constraint within the multinomial logit model ensures that the model coefficients can be considered as weights. Thus, the estimation results become interpretable with respect to the discriminative importance of the selected features, which is rated by a feature importance measure. In two application examples, data of a cell chip used for water quality monitoring experiments and phoneme data used for speech recognition, the interpretability as well as the selection results are examined. The classification performance is compared to various other classification approaches which are in common use.
\end{abstract}

\begin{keyword}
Functional Data Classification \sep Variable Selection \sep Feature Selection \sep Penalized Multinomial Logit Model \sep Nearest Neighbor Ensembles \sep Cell Chip Data \sep Phoneme Data
\end{keyword}

\end{frontmatter}

\section{Introduction}
Functional data analysis is an active field of research, and various methods for classification have been introduced by now. The term ``functional'' generally refers to the predictors used for classification, which are curves $x_i(t)$, with $t$ from a domain $\mathbb{D}$, typically an interval from $\mathbb{R}$. Functional covariates are infinite dimensional in theory, but, due to the limitations in data generation and processing, high dimensional in practice. In \cite{Ramsay2005}, a thorough introduction to functional data analysis and its most established modeling approaches can be found. Recent research on classification with functional predictors include \cite{James2001}, \cite{Mueller2005}, \cite{Rossi2006}, \cite{Epifanio2008}, \cite{Zhu2010}, \cite{Delaigle2012}, \cite{Gertheiss2013}, or \cite{Nguyen2016}.\\

\begin{figure}[]
\centering
\includegraphics[keepaspectratio,width=1\textwidth]{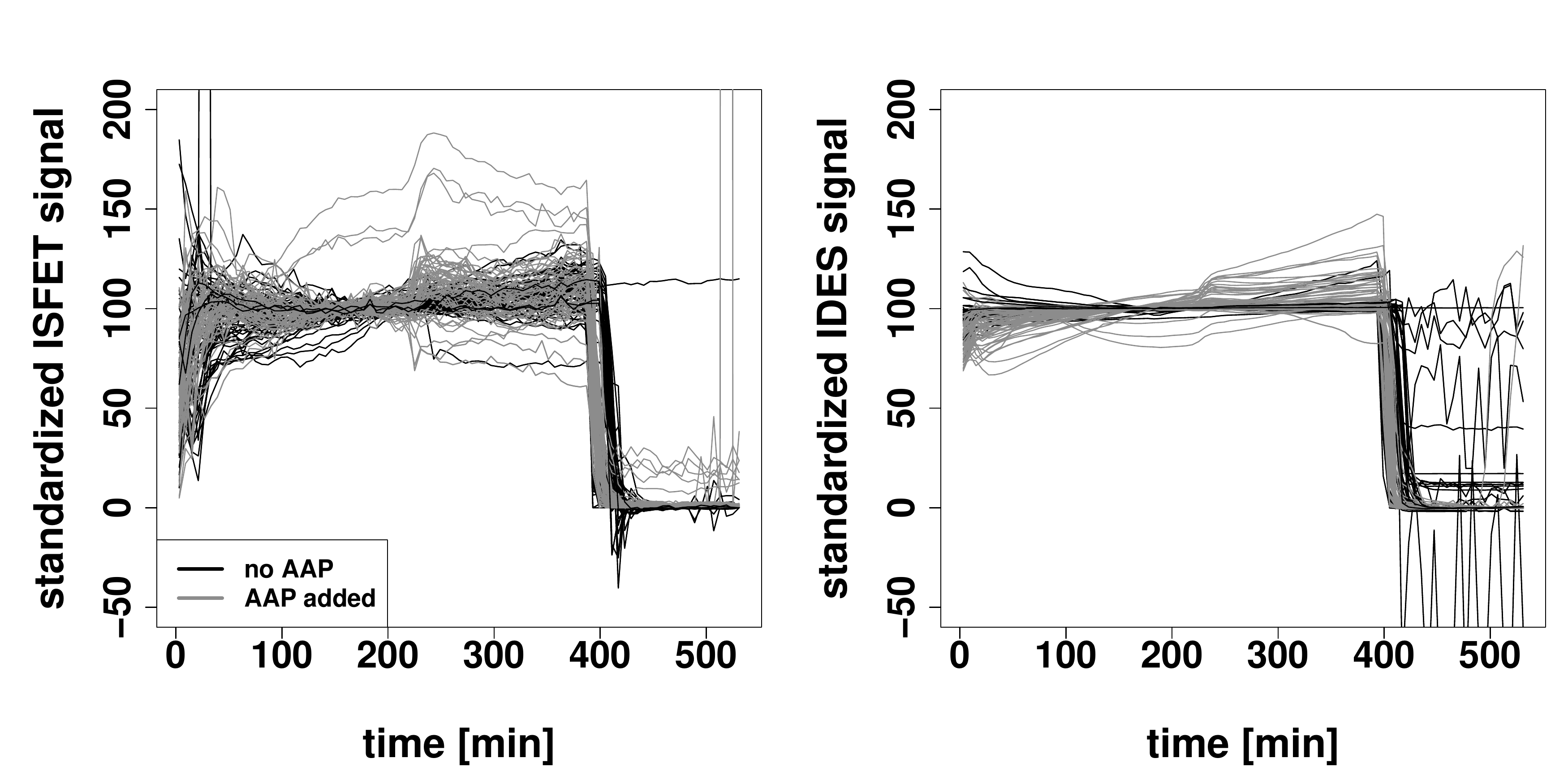}
\caption[Cell chip and cell chip data.]{\footnotesize{A total of $n=120$ standardized ISFET- and IDES-signals, recorded over time. Gray scales refer to the two classes of the discrimination task, with gray curves representing measurements with, and black curves measurements without the test substance AAP.}}
\label{fig:cellsignals}
\end{figure}

Examples for functional predictors are given by the data motivating our approach. These are the phoneme data introduced by \cite{Hastie1995} and cell chip data. The cell chip measurements are shown in Figure \ref{fig:cellsignals}. There are, among others, ISFET and IDES sensors implemented on a chips' surface, whose signals are recorded concurrently over time. Each sensor type captures aspects of the metabolism of a cell monolayer that is grown on the chips' surface, see also Section \ref{sec:appl} for further details. Since the signals represent physiological parameters of the cells' metabolism, and the latter is a continuous process over time, the sensor signals can be taken as realizations of a continuous function. 

The classification task in this data set results from disturbing the cells' habitual environment: While they are usually kept in nutrient medium, one can add test substances, as, for example, paracetamol (short: AAP), to this medium and monitor the cells' reactions. A functional two-class discrimination task is then given by comparing ISFET- and IDES-curves of measurements with (gray curves in Figure \ref{fig:cellsignals}) and without (black curves) adding AAP to the nutrient medium. Such discrimination tasks occur in the context of environmental monitoring, for example, water quality monitoring, where signals from clean samples are to be distinguished from polluted ones.\\

We introduce a novel interpretable feature selection method for classifying functional data, where the estimation of a functional $k$-nearest-neighbor ensemble ($k$NNE) is carried out by a penalized and constrained multinomial logit model (cMLM). The functional $k$NNE is set up by a large number of posterior probabilities for class membership that represent the ensemble members. Each posterior probability depends on the $k$ neighbors relative to the observation that is to be classified, as well as on a chosen semi-metric. As was shown by \cite{Ferraty2006}, semi-metrics are a suitable mathematical formulation to capture certain characteristics of a curve or function. Thus, by using adequate semi-metrics, a large variety of curve features, for example the curves' maxima or their mean values, can be included in the $k$NNE. The ensemble combines its members in a linear way, and each posterior probability, i.e.~ensemble member, is multiplied by an unknown coefficient which has to be determined. This yields the possibility to let members have differing importances. A (multinomial) logit link is then applied on these ensembles to obtain a classification model.\\
The ensemble coefficients are estimated by a cMLM, combined with a Lasso-type penalty. Lasso penalization allows for sparse results, estimating most coefficients equal to zero, as shown in \cite{Tibshirani1996}. Various Lasso-type penalties, suited for MLM, have been developed \citep[see][among others]{Argyriou2007, Simon2013, Chen2013, Vincent2014}. Apart from the standard global Lasso penalty \citep{Tibshirani1996}, we also employ the category-specific Lasso and the categorically structured (CATS) Lasso penalty \citep{Tutz2015b} for multi-class applications. The main advantage of using a Lasso-type penalty is the sparse estimation result enabling feature selection. Additionally we extend an ordinary penalized MLM such that a non-negativity constraint is put on the ensemble coefficients. This constraint ensures a proportional interpretability between the selected features with regard to the data background. As an example, consider two potentially interesting features in the above cell chip data, the step most curves exhibit around 210 minutes, and the distances between measurement curves in the region from about 210 to 400 minutes. By including corresponding semi-metrics in the $k$NNE and estimating the assigned coefficients by the penalized cMLM, one can decide whether the step or rather the curve distances yield more discriminative power.\\

As mentioned, several functional classification approaches have been developed in recent years. Some of them also use ensemble methods that combine scores or features of some kind in a linear combination. Among others, this kind of ensemble was used in \cite{Athitsos2005}, \cite{Preda2007}, or \cite{Araki2009}. Semi-metrics in the context of functional data classification were used for example in \cite{Ferraty2003} and \cite{Alonso2012}. All these approaches use a pre-defined and limited number of features that are given as input to a classification algorithm, and usually use the observed curve across its whole domain. This also holds for \cite{Matsui2014}, who use a functional logistic regression model to differ between functional observations of different classes, focusing on variable selection. In contrast, our approach can handle a very large number of semi-metrics, i.e.~features, in the functional $k$NNE. This includes taking only specific parts of the curves into account, which might be more significant in certain data situations than using the whole curve, as the above cell chip data suggests. Additionally, for most functional classification methods, the extension to multiple functional covariates is not straight forward, and the estimation results are not interpretable with regard to the data background. Interpretability also is the most weak point in the method by \cite{Moeller2016}, who built random forests out of summary quantities calculated from (randomly chosen) curve segments. The approach presented here offers both, the potential inclusion of multiple functional (and non-functional) covariates as well as interpretability. These two advantages are also present in the method of \cite{Fuchs2015}, who apply a similar ensemble on functional covariates, but estimate the ensemble coefficients by means of Brier score minimization. This implies a more restrictive constraint on the coefficients than is necessary with the estimation via a penalized cMLM. Further advantages that arise from the use of a MLM are that MLM can readily be adapted to ordinal data, and estimation is fast and stable. The penalty might be chosen with respect to probable interrelations between the coefficients, such as the CATS Lasso penalty.\\

The remainder of this manuscript is organized as follows: In Section \ref{sec:method}, our approach is introduced in detail. In Sections \ref{sec:appl} and \ref{sec:appl_phonemes}, the performance of the presented approach is analysed with respect to the cell chip and phoneme data sets, and is compared to other (functional) classification approaches. The manuscript closes with a discussion of the results and further possible extensions concerning our method in Section \ref{sec:discuss}.\\
We provide exemplarily code as well as the respective data to make the application results fully reproducible.


\section{Method}
\label{sec:method}
In this section, the proposed approach is given in detail. First, the functional $k$NN ensemble setup is described, with a focus on the basic design and the semi-metrics that capture specific curve features. The following Section \ref{sec:method_MLM} introduces penalized cMLM and various Lasso-type penalties. Section \ref{sec:estimateComputation} outlines details of the estimation of our penalized cMLM model. In Section \ref{sec:CompMeths}, a short description of alternative classification techniques is given, and the Brier score and misclassification rate are defined as performance measures to explore the predictive capability of the single methods. An additional feature importance measure is introduced, allowing for the proportional interpretability of our method's selection results.


\subsection{Functional nearest neighbor ensembles}
\label{sec:method_fknn}
Let $x_i(t)$, $i=1,\ldots n$, be a set of curves, i.e.~functional predictors, with $t$ from a domain $\mathbb{D}\subseteq\mathbb{R}$. Assume that each curve belongs to one of $G$ different classes, denoted by $y_i\in\{1,\ldots G\}$. Further, let $x_i^{(a)}(t)$ denote the $a$th derivative of the functional covariate $x_i(t)$, and $d\left(x_i^{(a)}(t),x_j^{(a)}(t)\right)$ a semi-metric of (the derivatives of) two functional covariates $x_i^{(a)}(t)$, $x_j^{(a)}(t)$, $i\neq j$.


\subsubsection{Translating curve features into probabilities via semi-metrics}
A simple functional $k$-nearest-neighbor ensemble can be specified through a set of various semi-metrics $d_l(\cdot,\cdot)$, where $l=1,\ldots p$ is the index for the different semi-metrics. Each semi-metric is chosen such that it extracts a particular feature from the curves. Apart from the semi-metrics that were already mentioned in the Introduction, representing the step or distances in the cell chip data, further examples are the covariates' maxima or curvatures. Since some curve characteristics might be amplified after derivation, we use the derivatives of the functional predictors as well as the original curves. Thus, the nearest neighbor ensemble to be considered uses semi-metrics $d_l\bigl(x_i^{(a)}(t),x_j^{(a)}(t)\bigr)$, applied on the quantitites $x^{(a)}_i(t)$, $x^{(a)}_j(t)$, $i\neq j$. The whole set of semi-metrics we use is given in Table \ref{tab:semmets}.

\begin{sidewaystable}
\footnotesize
\centering
	\begin{tabular}{p{3.5cm}p{6.5cm}p{8cm}}
	\toprule
	Semi-metric & Formula & The semi-metric focuses on...\\
	\midrule
	$d^{Eucl}(x^{(a)}_i(t),x^{(a)}_j(t))$ & $\sqrt{\int_\mathbb{D}\left(x_i^{(a)}(t)- x_j^{(a)}(t)\right)^2 dt}$ & 
		... the absolute distance of two curves (or their derivatives).\\
	\midrule
	$d_{\tau}^{Scan}(x^{(a)}_i(t),x^{(a)}_j(t))$ & $\sqrt{\int_\mathbb{D}\left(\phi_{\tau}(t)\left(x_i^{(a)}(t)- x_j^{(a)}(t)\right)\right)^2 dt}$, $\qquad\qquad$ with $\tau\in\mathbb{D}$ &
	... the absolute distances of weighted profiles of the original curves (or their derivatives), centered around $\tau$. \\
	\midrule
	$d_{\mathbb{D}_{small}}^{shortEucl}(x^{(a)}_i(t),x^{(a)}_j(t))$ &  $\sqrt{\int_{\mathbb{D}_{small}}\left(x_i^{(a)}(t)- x_j^{(a)}(t)\right)^2 dt}$ &
	... the absolute distance on a limited part of the domain of definition $\mathbb{D}_{small}\subset\mathbb{D}$ of two curves (or their derivatives).\\
	\midrule
	$d^{Mean}(x^{(a)}_i(t),x^{(a)}_j(t))$ &  $\left|\int_\mathbb{D}x_i^{(a)}(t)dt-\int_\mathbb{D}x_j^{(a)}(t)dt\right|$
	&
	... the similarity of mean values of the whole curves (or their derivatives).\\
	\midrule
	$d^{relAreas}(x^{(a)}_i(t),x^{(a)}_j(t))$ &  $\Bigg|\left|\frac{\int_{\mathbb{D}_1}x_i^{(a)}(t)dt} {\int_{\mathbb{D}_2}x_i^{(a)}(t)dt}\right|-$
	$ \left|\frac{\int_{\mathbb{D}_1}x_j^{(a)}(t)dt} {\int_{\mathbb{D}_2}x_j^{(a)}(t)dt}\right|\Bigg|$ 
	&
	... the similarity of the relation of areas on parts of the domain of definition, $\mathbb{D}_1$, $\mathbb{D}_2\subset\mathbb{D}$.\\
	\midrule
	$d_{bo}^{Jump}(x_i(t),x_j(t))$ &  $\Big|\left(x_i(t_b)-x_i(t_o)\right)-$ 	$\left(x_j(t_b)-x_j(t_o)\right)\Big|$
	&
	... the similarity of jump heights at points $t_b$, $t_o\in\mathbb{D}$.\\
	\midrule
	$d^{Max}(x^{(a)}_i(t),x^{(a)}_j(t))$ &  $\left|\text{max}\left(x_i^{(a)}(t)\right)- \text{max}\left(x_j^{(a)}(t)\right)\right|$
	&
	... the difference of the curves' (or their derivatives') global maxima.\\
	\midrule
	$d^{Min}(x^{(a)}_i(t),x^{(a)}_j(t))$ &  $\left|\text{min}\left(x_i^{(a)}(t)\right)- \text{min}\left(x_j^{(a)}(t)\right)\right|$
	&
	... the difference of the curves' (or their derivatives') global minima.\\
	\midrule
	$d^{Points}(x^{(a)}_i(t),x^{(a)}_j(t))$ & $\frac{1}{E}\sum_{e=1}^E \left|x_i^{(a)}(t_e)- x_j^{(a)}(t_e)\right|$
	&
	... the differences at certain observation points $t_e$ (also called ``points of impact").\\
	\bottomrule
	\end{tabular}
\caption{\footnotesize{Semi-metrics used to set up the $k$-nearest-neighbor ensemble.}}
\label{tab:semmets}
\end{sidewaystable}

After having defined adequate semi-metrics, possibly with respect to expert knowledge about the data at hand, the corresponding semi-metric dependent neighborhoods can be defined. With respect to a generic or a new observation $(y^*,x^{*}(t))$ and a specific semi-metric $d_{l}(\cdot,\cdot)$, the learning sample $(y_i,x^{(a)}_i(t))$, $i=1,\ldots n$ is ordered such that
\begin{center}
$d_{l}\big(x^{*(a)}(t),x^{(a)}_{(1)}(t)\big) \leq \ldots\leq d_{l}\big(x^{*(a)}(t),x^{(a)}_{(k)}(t)\big) \leq \ldots\leq d_{l}\big(x^{*(a)}(t),x^{(a)}_{(n)}(t)\big).$\nonumber
\end{center}
Hence, the $x_{(i)}^{(a)}(t)$ are the curves (or their derivatives) from the learning sample, ordered by their distance to $x^{*(a)}(t)$ as measured by the semi-metric $d_l(\cdot,\cdot)$, so that one can define a neighborhood of the $k$ nearest neighbors of $x^{*(a)}(t)$ by
\begin{center}
$\mathcal{N}^k_{l}\bigl(x^{*(a)}(t)\bigr)=\bigl\{x^{(a)}_{j}(t):\ d_{l}(x^{*(a)}(t),x^{(a)}_{j}(t))\leq d_{l}(x^{*(a)}(t),x^{(a)}_{(k)}(t))\bigr\}$.
\end{center}
Let 1$(\cdot)$ denote the indicator function. Then a single ensemble member is given by the estimated posterior probability $w_{gl}$ that $x^{*(a)}(t)$ is from class $g$,
\begin{eqnarray}
w_{gl}=\frac{1}{k}\sum_{\left\{j:\ x^{(a)}_j(t)\in \mathcal{N}^k_{l}\bigl(x^{*(a)}(t)\bigr)\right\}}1(y_j=g).\nonumber
\end{eqnarray}
It is determined by the number of nearest neighbors $k$, the semi-metric $d_l(\cdot,\cdot)$, and the order of the derivative $a$. For a specific observation $(y_i,x_i^{(a)}(t))$ from the learning sample, the posterior probabilities are obtained by a leave-one-out procedure, which is formally given by 
\begin{eqnarray}
w_{igl}=\frac{1}{k}\sum_{\left\{j\neq i:\ x^{(a)}_j(t)\in \mathcal{N}^{k+1}_{l}\bigl(x_i^{(a)}(t)\bigr)\right\}}1(y_j=g), \qquad\qquad i=1,\ldots,n.\nonumber
\end{eqnarray}


\subsubsection{Simple functional nearest neighbor ensembles}
The observed classes $y_i$ from the previous section are considered as realizations of random variables $Y_i$ that take on values in $\{1,\ldots,G\}$. Since the methodology uses $p$ semi-metrics, it comprises $p$ posterior probabilities $w_{igl}$ for each observation $i$ and class $g$. These single posterior probabilities can be combined in a linear combination to obtain a simple ensemble model for the overall posteriori probability that observation $x_i(t)$ is from class $g$, given by
\begin{align}
\pi_{ig} &=\sum_{l=1}^{p} w_{igl}c_l, \label{eq:simpleensemble}\\
\intertext{where $c_l$ are weights that have to be estimated and must satisfy}
 \sum_{l=1}^p c_l& = 1 \quad \text{and} \quad c_l \ge 0\ \forall l. \label{eq:ensemblerestriction} 
\end{align}
Similar to the $k$-nearest-neighbor classifier for multivariate data, each observation is assigned to the class of highest posterior probability, i.e.~$\hat{y}_i=\underset{g}{\text{max}}\left(\pi_{ig}\right)$.\\

In addition to the various semi-metrics used in Ensemble (\ref{eq:simpleensemble}), we also use different sizes for the number of nearest neighbors $k\in\mathcal{K}_{nN}=\{k_1,\ldots k_M\}$ as well as varying orders of derivation $a\in\{a_1,\ldots a_O\}$. The definitions of the above neighborhood $\mathcal{N}_{l}^k\left(x^{*(a)}(t)\right)$ and single posterior probability estimates $w_{igl}$ are adapted accordingly. This means that from now on, the index $l$ does not refer to a single semi-metric, but represents an ensemble member determined by an unique tuple $\{d(\cdot,\cdot),a,k\}$ of a specific semi-metric $d(\cdot,\cdot)$, a number of nearest neighbors $k$ and an order of derivative $a$.\\
If multiple functional covariates are observed, it might be necessary to include covariate type-specific semi-metrics into the ensemble. For instance this might arise if the covariate types originate from different domains, as for example time and wavelengths. Another situation is given in our cell chip application in Section \ref{sec:appl}, where two different sensors measure different physiological parameters. With $R$ functional predictor types, one has observations $\left(y_i, x_{i1}(t),\ldots,x_{ir}(t)\right)$, $r=1,\ldots R$, such that each neighborhood, and with that each ensemble member and ensemble coefficient, additionally depends on the covariate type. \\
Let $q$, $R$, $M$, and $O$ denote the numbers of semi-metrics, covariate types, nearest neighbors, and orders of derivation used. Then, the ensemble comprises a total number of $p=q\cdot R\cdot M\cdot O$ members. \\

One way to estimate Ensemble (\ref{eq:simpleensemble}) with respect to Constraint (\ref{eq:ensemblerestriction}) is to optimize some loss function like the Brier score. Such an estimation approach does not allow for category-specific ensemble coefficients, and the extension to ordinal classes is not self-evident. In the next Section, we propose an alternative estimation technique.


\subsection{The penalized and constrained multinomial logit model}
\label{sec:method_MLM}
Alternatively to loss functions, the estimation of Ensemble (\ref{eq:simpleensemble}) can be performed via a multinomial logit model, yielding sparse and interpretable results if being penalized and constrained adequately. To illustrate this, for $g=1,\ldots G-1$, let $v_{igl} = (w_{igl} - w_{iGl})$ denote the differences in posterior probability between classes. For a more compact notation, let $\bmath{v}_{ig} = (v_{ig1},\ldots,v_{igp})^T$ and $\bmath{v}_i = (\bmath{v}_{i1}^T,\ldots,\bmath{v}_{i,G-1}^T)^T$.
We consider, for $g=1,\ldots G-1$, the following constrained multinomial logit model:
\begin{eqnarray}\label{eq:cmnl}
P(Y_i = g|\bmath{v}_i) &=& \pi_{ig} \nonumber\\
&=& \frac{\exp(\bmath{v}_{ig}^T\bmath{c})}{1 +\sum_{s=1}^{G-1} \exp(\bmath{v}_{is}^T\bmath{c})} \nonumber\\
&=& \frac{\exp\Bigl(\sum_{l=1}^{p} v_{igl}c_l\Bigr)}{1+\sum_{s=1}^{G-1} \exp\Bigl(\sum_{l=1}^{p} v_{isl}c_l\Bigr)} \qquad \text{s.t. $c_l \ge 0 \ \forall l.$}
\end{eqnarray}
The probability for the ``reference class'' $G$ is trivially given by
\[\pi_{iG} = 1-\sum_{s=1}^{G-1}\pi_{is} = \frac{1}{1 +\sum_{s=1}^{G-1} \exp(\bmath{v}_{is}^T\bmath{c})}.\]\\
The novelty in Model (\ref{eq:cmnl}) relative to an ordinary MLM is the constraint on the model coefficients. The restriction to values equal or above zero allows a proportional interpretation of the coefficients' values with respect to the data background, and thus yields a relevant improvement.\\

Model \eqref{eq:cmnl} can also be rewritten in terms of log odds with regard to the reference class $G$ as follows:
\begin{eqnarray}\label{eq:cmnlodds}
\log\left(\frac{P(Y_i = g|\bmath{v}_i)}{P(Y_i = G|\bmath{v}_i)}\right) &=& \bmath{v}_{ig}^T\bmath{c} = \sum_{l=1}^{p} v_{igl}c_l \nonumber\\
&=&\sum_{l=1}^{p} (w_{igl}-w_{iGl})c_l \qquad \text{s.t. $c_l \ge 0 \ \forall l$}\nonumber
\end{eqnarray}
for $g=1,\ldots G-1.$
If $w_{igl} > w_{iGl}$ with $c_l \neq 0$, then the logit will change in favor of class $g$ over $G$ accordingly, and vice versa for $w_{igl} < w_{iGl}$. Hence, when classes show differences in their posterior probabilities which are based on a particular curve feature $l$, this information gets automatically translated into a change of the overall class probability. 
Given the difference building on the posterior probabilities, which in our model have the role of covariates, the nonnegativity constraint that is imposed on the parameters $c_l$ allows to interpret them as weights that reflect the importance of the different curve features for the classification. From a more technical point of view, the $w_{igl}$ are class-specific covariates, and using their difference to a reference class (here $G$) is necessary to make the MLM identifiable. The use of these differences also arise naturally when the MLM is motivated via latent utility maximization as shown in \citet{Mcfadden73}. 

Since the linear predictors $\bmath{v}_{ig}^T\bmath{c}$ are transformed to class probabilities via the multinomial logit link, feasible estimates $\hat{\pi}_{ig} \in [0,1]$ are guaranteed for any parameter estimate $\hat{\bmath{c}}$. This holds for both the observed data as well as in prediction, so that Model \eqref{eq:cmnl} can easily be extended to more flexible, class-specific weights $c_{gl}$:
\begin{eqnarray}\label{eq:cmnl2}
P(Y_i = g|\bmath{v}_i) &=& \frac{\exp(\bmath{v}_{ig}^T\bmath{c}_g)}{1 +\sum_{s=1}^{G-1} \exp(\bmath{v}_{is}^T\bmath{c}_s)} \nonumber\\
&=& \frac{\exp\Bigl(\sum_{l=1}^{p} v_{igl}c_{gl}\Bigr)}{1+\sum_{s=1}^{G-1} \exp\Bigl(\sum_{l=1}^{p} v_{isl}c_{sl}\Bigr)} \qquad \text{s.t. $c_{gl} \ge 0 \ \forall g,l.$}
\end{eqnarray}
It was shown in \cite{Gertheiss2009} that $\hat{\pi}_{ig} \in [0,1]$ cannot be guaranteed in prediction if one uses class-specific weights $c_{gl}$ in the simple linear ensemble approach \eqref{eq:simpleensemble}. Hence, the modeling option \eqref{eq:cmnl2} is a major advantage of our cMLM approach since it allows a more flexible model for classification tasks in problems with more than two classes. 
The analysis of the phoneme data in Section \ref{sec:appl_phonemes} illustrates this advantage on real data.


\subsubsection{Penalization for the constrained MLM}
\label{sec:method_MLM_penals}
Depending on the data at hand and the choices of the sets of predictors, semi-metrics, the number of different neighborhood sizes and the number of different orders of covariate derivatives used, the number $p$ of coefficients that has to be estimated can easily reach orders of $10^2$ to $10^5$. To regularize the estimates and to find an interpretable set of curve features that explains the curves' class memberships, we propose to use penalized estimation with variable selection penalties. 

In the following, let $\log(L(\bmath{c}))$ denote the log-likelihood of Models \eqref{eq:cmnl} or \eqref{eq:cmnl2}, let $J(\bmath{c})$ denote a penalty term and let $\lambda \ge 0$ denote a tuning parameter that controls the strength of the penalization. 
For the model from \eqref{eq:cmnl} that uses global weights on all curve features, a suitable penalty is given by the Lasso of \citet{Tibshirani1996}, yielding 
\begin{equation}\label{eq:globallasso}
\hat{\bmath{c}} = \underset{\bmath{c}}{\text{argmax}}\,\left(\log(L(\bmath{c})) \ - \ \lambda J(\bmath{c}) \ \, =\ \,  \log(L(\bmath{c})) \ -\ \lambda \sum_{l=1}^p |c_l|  \qquad \text{s.t. $c_l \ge 0 \ \forall l$}\right).
\end{equation}
Due to its mathematical properties, the Lasso penalty $J(\bmath{c}) = \sum_{l=1}^p |c_l|$ induces sparse estimates for suitably chosen $\lambda$, that is, one obtains $\hat{c}_l = 0$ for many $l$, yielding selection of curve features. 

For the model from \eqref{eq:cmnl2} with class-specific weights, the Lasso estimates are given by
\begin{equation}\label{eq:classwiselasso}
\hat{\bmath{c}} = \underset{\boldsymbol{c}}{\text{argmax}}\,\left( \log(L(\bmath{c})) \ -\ \lambda \sum_{l=1}^p\sum_{g=1}^{G-1} |c_{gl}|  \qquad \text{s.t. $c_{gl} \ge 0 \ \forall g,l$}\right).
\end{equation}
In this case, the Lasso approach induces solutions that are sparse on the parameter level, i.e.~it is possible to obtain, for example, $\hat{c}_{1l} > 0$ and $\hat{c}_{2l} = 0$. Since the denominator in \eqref{eq:cmnl2} is influenced by all parameters, the respective probability $P(Y_i = 2 |\bmath{v}_i)$ would still be influenced by curve feature $l$. 

To obtain a proper selection of curve features, one can adopt the Categorically Structured Lasso approach of \citet{Tutz2015b}, which computes estimates according to
\begin{equation}\label{eq:cats}
\hat{\bmath{c}} = \underset{\boldsymbol{c}}{\text{argmax}}\,\left( \log(L(\bmath{c})) \ -\ \lambda \sqrt{G\!-\!1} \sum_{l=1}^p \sqrt{\sum_{g=1}^{G-1}c_{gl}^2}  \qquad \text{s.t. $c_{gl} \ge 0 \ \forall g,l$}\right).
\end{equation}
The CATS approach treats all parameters that belong to the same curve feature $l$ as one parameter group that is removed from the model jointly. However, here the CATS approach can produce solutions with so-called `within-group-sparsity', which is in stark contrast to the behavior of CATS in the context of \citet{Tutz2015b}. This phenomenom has technical reasons, see also Equation \eqref{eq:proxopcatsconstr} in the following section.


\subsection{Computation of estimates}
\label{sec:estimateComputation}
From a technical point of view, our constrained and penalized MLM is simply an ordinary penalized MLM with class-specific covariates and a constraint. Therefore, our model is covered by the framework of \citet{Tutz2015b} except for the nonnegativity constraint on the parameters. Hence, the FISTA algorithm for the estimation of a penalized MLM proposed in \citet{Tutz2015b} must be adapted to incorporate this constraint. The core problem to solve is the so-called proximal operator with nonnegativity constraint. For the Lasso penalty on global parameters as in \eqref{eq:globallasso}, and with $\bmath{u} \in \mathbb{R}^p$ denoting an arbitrary and generic input vector, this problem has the following form:
\begin{equation*}\label{eq:proxoplasso}
\bmath{Prox}_{\text{lasso}} (\bmath{u}|\; \lambda) \ =\  \underset{\boldsymbol{c}\in \mathbb{R}_{\geq 0}^p}{\text{argmin}}\,\left( \tfrac{1}{2} ||\bmath{c} - \bmath{u}||_2^2 + \lambda \sum_{l=1}^p|c_l|\right).
\end{equation*}
As proven in \citet{Jenatton2011}, this problem is solved by simply replacing all negative entries of the input vector with zero, which reduces the problem to the well-studied unconstrained proximal operator. With $[\bmath{u}]_{+} = \max(\bmath{u},0)$ (to be understood entry-wise), one obtains
\begin{align*}\label{eq:proxoplassoconstr}
\underset{\boldsymbol{c}\in \mathbb{R}_{\geq 0}^p}{\text{argmin}}\,\left( \tfrac{1}{2} ||\bmath{c} - \bmath{u}||_2^2 + \lambda \sum_{l=1}^p|c_l|\right) \ &=\  \underset{\boldsymbol{c}\in \mathbb{R}^p}{\text{argmin}}\,\left( \tfrac{1}{2} ||\bmath{c} - [\bmath{u}]_{+}||_2^2 + \lambda \sum_{l=1}^p|c_l|\right)\\
=\ \left(\Bigl[[u_l]_{+}-\lambda\Bigr]_{+}\right)_{l=1,\ldots p}\ &=\ \left(\max\Bigl(\max(u_l,0)-\lambda,\  0\Bigr)\right)_{l=1,\ldots p}.
\end{align*}
An equivalent statement holds for the Lasso penalty on class-specific weights from \eqref{eq:classwiselasso}. For the CATS penalty from \eqref{eq:cats}, one obtains for $l=1\ldots p$
\begin{equation}\label{eq:proxopcatsconstr}
\underset{\boldsymbol{c}_l\in \mathbb{R}_{\geq 0}^{G-1}}{\text{argmin}}\,\left( \tfrac{1}{2} ||\bmath{c}_l - \bmath{u}_l||_2^2 + \lambda \sqrt{G\!-\!1} \sqrt{\sum_{g=1}^{G-1}c_{gl}^2}\right) \ =\  \left[1 - \frac{\lambda\sqrt{G\!-\!1}}{\bigl|\bigl|[\bmath{u}_l]_{+}\bigr|\bigr|_2}\right]_{+} [\bmath{u}_l]_{+}.
\end{equation}

The corresponding estimation algorithm is implemented in the publicly available \texttt{R}-package \texttt{MRSP} \citep{MRSP2015}. The nonnegativity constraint on the parameters can be activated by using the argument ``\texttt{nonneg = TRUE}''.\\


\subsection{Competing methods and prediction performance measures}
\label{sec:CompMeths}
To be able to evaluate the prediction performance of the $k$NNE estimated via penalized cMLM, we compare all results to alternative approaches. Since only few functional classification methods introduced so far come with an implementation, we will also use some multivariate discrimination techniques that are known for their good performance. To this end, we either use the discretized covariates if appropriate, or otherwise compute a functional principal component (FPC) analysis and use the respective scores as inputs. To the best of our knowledge, the choice of the number of components in FPC analysis is still an open problem. An approach used by some researchers \citep[e.g.][]{Hall2001} is to use those scores that explain a specified percentage of the sample variability. We do the same, choosing the number of scores such that at least 95\% variability is explained. The FPC computation is carried out with the \texttt{fpca.sc}-function of the \texttt{R}-package \texttt{refund} \citep{Di2009, refund2013, Goldsmith2013}.\\
In the following, all methods included in the comparison are shortly presented. For details on the single methods, the reader is referred to the literature. The calculations for all methods, including the penalized cMLM, are carried out using the software environment \texttt{R} \citep{RCore2016} and respective add-on packages. Details on the latter as well as on the choices of parameters, where appropriate, are given in the following. If not stated otherwise, the default parameters are used.

\begin{description}
\item[\textit{Penalized constrained multinomial logit model}] (\textit{cMLM}) The method introduced in Sections \ref{sec:method_fknn} - \ref{sec:estimateComputation}. Before modeling, both the learning as well as the test data set are normalized to a standard deviation of one. The penalty parameter $\lambda$ is chosen from a predefined grid of values. The model choice bases on the minimization of the mean Akaike information criterion (AIC). The abbreviation \textit{cMLM} is used whenever a global Lasso penalty is employed. The abbreviation \textit{cs cMLM} implies that a category-specific Lasso penalty is used, \textit{csCATS cMLM} denotes the usage of a categorically structured Lasso penalty. For computation, the add-on package \texttt{MRSP} \citep{MRSP2015} is used.

\item[\textit{$k$-nearest-neighbor ensemble}] \citep[\textit{kNN Ensemble}, ][]{Fuchs2015} The ensemble is, apart from the standard deviation's normalization, identical to the one set up for the cMLM, i.e.~Model (\ref{eq:simpleensemble}) has to satisfy Constraints (\ref{eq:ensemblerestriction}). The ensemble coefficients are estimated via the Brier score minimization used in the original research article. Software from the respective supplement is used.

\item[\textit{Nonparametric functional classification}] \citep[\textit{NPFC}, ][]{Ferraty2003} We use four semi-metrics the approach offers, namely the Euclidian distance of the $a$th derivative of the functional covariates (abbreviation \textit{NPFC - deriv}), the Euclidian distance of the Fourier expansions of the functional covariates (abbreviation \textit{NPFC - Fourier}), the multivariate partial least squares regression semi-metric (abbreviation \textit{NPFC - mplsr}), and the functional principal component semi-metric (abbreviation \textit{NPFC - pca}). For details on the single semi-metrics, we refer to \cite{Ferraty2006}. All semi-metrics require the choice of at least one parameter. These are chosen by minimizing the mean misclassification error of a 10-fold CV. The modeling software can be found at \textit{http://www.math.univ-toulouse.fr/~ferraty/SOFTWARES/NPFDA/index.html}.

\item[\textit{Functional linear model}] \citep[\textit{FLM - log}, ][]{Ramsay2005} In the case of a two-class problem, we use a parametric functional model, which is implemented in the \texttt{gam}-function of the \texttt{R} package \texttt{mgcv} \citep{mgcv2014}. The number of basis functions used for each smooth term has to be chosen. We test a grid $\{3,\ldots 10\}$ and choose the number with minimal mean misclassification rate in the test data.

\item[\textit{Functional random forests}] \citep[\textit{fRF}, ][]{Moeller2016} The software has kindly been provided by the authors of \cite{Moeller2016}. The classification is carried out by the function \texttt{FuncRandomForest}. Differing from the default call, the variables \texttt{importance} and \texttt{overlap} are set to \texttt{TRUE} to have access to the variable importance measure of the method and achieve a more flexible interval choice. The model parameters $\lambda$ and $c$ were each chosen from predefined grids via minimizing the mean misclassification rate of a 5-fold CV.

\item[\textit{Linear discriminant analysis}] \citep[\textit{LDA}, ][]{Fisher1936, Rao1973} We apply \textit{LDA} to the FPC scores, as done by \cite{Ramsay2002}. \textit{LDA} is implemented in the \texttt{R} package \texttt{MASS} \citep{mass2014}, function \texttt{lda}.

\item[\textit{Penalized discriminant analysis}] \citep[\textit{PDA - cov., }][]{Hastie1995} \textit{PDA} was especially designed for high-dimensional and highly correlated covariates \citep{Hastie1995}, such that it can be applied on the discretized data. The approach is implemented in the \texttt{R} package \texttt{mda} \citep{mda2015}, function \texttt{fda}.

\item[\textit{Multinomial model}] \citep[\textit{mM}, see e.g.][]{Tutz2012} A multinomial logistic regression model is used on the FPC scores. This method is implemented in the \texttt{maxent}-function of the \texttt{R} package \texttt{maxent} \citep{maxent2013}.

\item[\textit{Support vector machines}] \citep[\textit{SVM}, ][]{Vapnik1996} We use the implementation of the \texttt{R} package \texttt{e1071} \citep{e10712014} by using the function \texttt{svm}, with \texttt{probability=TRUE} and default settings else. The \textit{SVM} is applied to both, the discretized data, referred to by \textit{SVM - cov.}, and the FPC scores (\textit{SVM - FPCs}).

\item[\textit{Random forests}] \citep[\textit{RF}, ][]{Breiman2001} The method is implemented in the \texttt{R} package \\
\texttt{randomForest} \citep{rf2012}. We used the \texttt{randomForest}-function. \textit{RF} are applied to both, the discretized data (\textit{RF - cov.}) and the FPC scores (\textit{RF - FPCs}).

\item[\textit{Regularized discriminant analysis}] \citep[\textit{RDA}, ][]{Guo2007} \textit{RDA} penalizes a \textit{LDA} and was designed for high-dimensional data, thus being applied to the discretized covariates. It is available from the \texttt{R} package \texttt{rda} \citep{rda2012}, function \texttt{rda}.

\item[\textit{Sparse discriminant analysis}] \citep[\textit{SDA}, ][]{Clemmensen2011} Another modification of \textit{LDA} is \textit{SDA}, which we apply to the discretized covariates. An implementation can be found in the \texttt{R} package \texttt{sda} \citep{sda2015}, function \texttt{sda}.
\end{description}

To be able to compare the prediction performances of the competing methods described above, we will use two performance measures. The first is the Brier score operating on the coded response $z_{ig}=1$ if $y_{i}=g$ and $z_{ig}=0$ otherwise,
\begin{eqnarray}
Q&=&\frac{1}{n_{test}}\frac{1}{G}\sum_{i=1}^{n_{test}} \sum_{g=1}^G \left(z_{ig}-\pi_{ig}\right)^2, \nonumber
\label{eq:glBrier}
\end{eqnarray}
introduced by \cite{Brier1950}. $n_{test}$ denotes the sample size of the test data. As shown by \cite{Gneiting2007}, the Brier score is a proper scoring rule. Moreover, it is the only one that fulfills, up to a positive linear transformation, the properties that \cite{Selten1998} demanded of scoring rules. This implies that, in contrast to, for example, the logarithmic score, the Brier score does not react strongly on small differences between small probabilities, especially probabilities of values (around) zero. Although improper in the above sense, the second performance measure we use is the classical misclassification rate (MCR)
\begin{equation}
MCR=\frac{1}{n_{test}}\sum_{i=1}^{n_{test}} 1\left(y_{i}=\hat{y}_{i}\right).\nonumber
\end{equation}
Here, 1$(\cdot)$ again denotes the indicator function, $y_{i}$ denotes the true class of observation $x_{i}(t)$, and $\hat{y}_{i}$ the class assigned by the method considered.\\

In addition to these performance measures we introduce a score that indicates how important the coefficients, estimated by the penalized cMLM, are relative to each other, i.e.~which of the features corresponding to the coefficients yields most discriminative power. The score is called the relative feature importance (RFI) measure and yields the percental importance per estimated coefficient $\hat{c}_l$. It is defined by
\begin{equation}
RFI_l=100\cdot \frac{\sum_{g=1}^G\hat{c}_{gl}}{\sum_{g=1}^G\sum_{s=1}^p \hat{c}_{gs}}.\nonumber
\end{equation}
In the case of global coefficients, the RFI measure simplifies to $RFI_l=100\cdot\frac{\hat{c}_{l}}{\sum_{s=1}^p \hat{c}_{s}}$.


\section{Application to real world data - cell based sensor chips}
\label{sec:appl}
\begin{figure}[]
\centering
\includegraphics[keepaspectratio,width=1\textwidth]{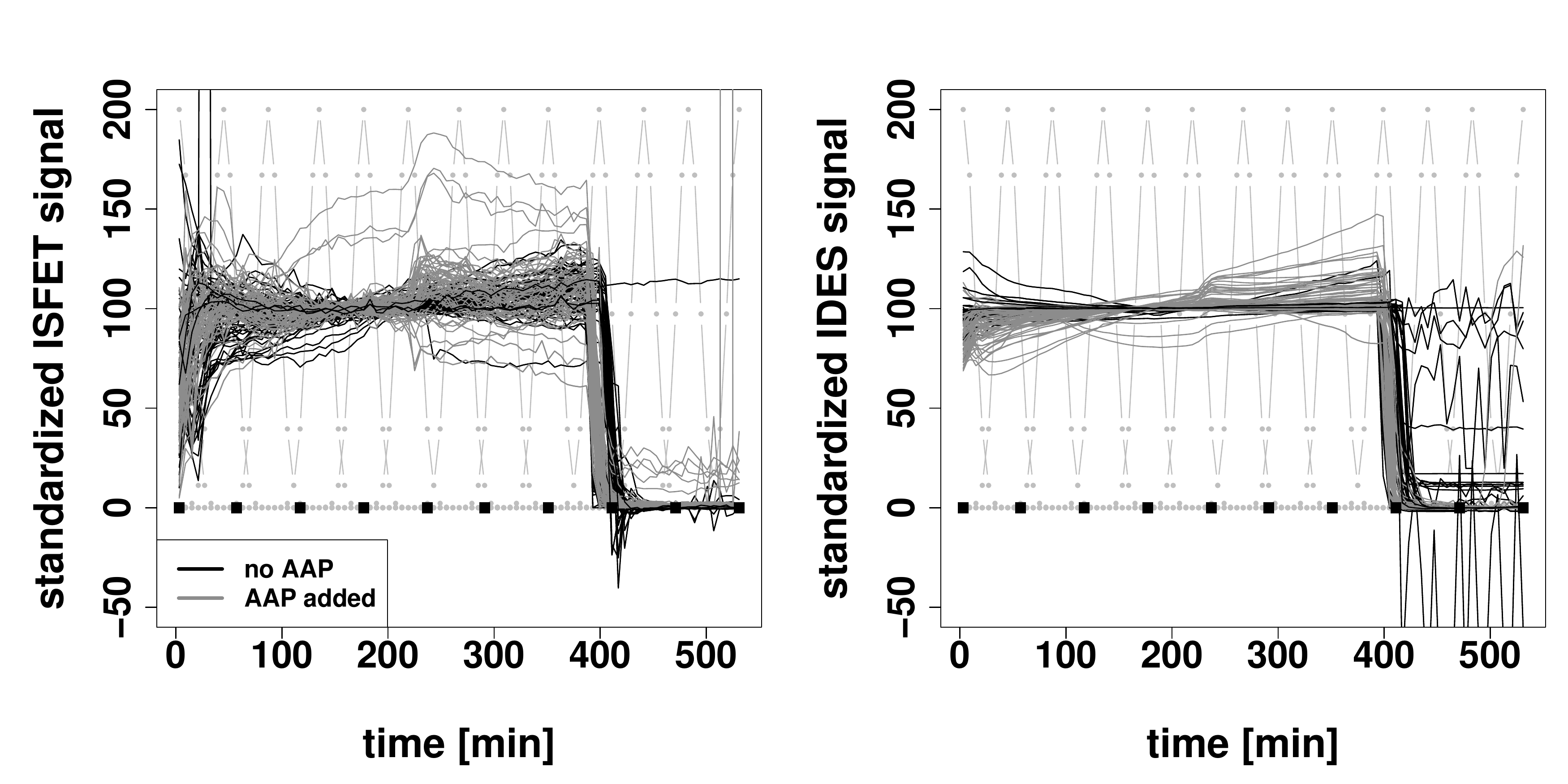}
\caption[Cell chip and cell chip data.]{\footnotesize{The same $n=120$ standardized ISFET- and IDES-signals as before. The light gray, dotted lines depict the function $\phi_{\tau}(t)$ in $d_{\tau}^{Scan}$ used at certain observation points $\tau$, the impact points $t_e$ used in $d^{Points}$ are depicted as black boxes.}}
\label{fig:cellsignals_params}
\end{figure}

Cell based sensor technologies attract much interest in biomechanical engineering, especially in application fields concerned with environmental quality monitoring, see \cite{Bohrn2012} or \cite{Kubisch2012} for some examples and references.\\
In this study we use cell based chips, which means that the chips' surfaces are covered with a monolayer of a living cell population. There are different kinds of sensors distributed across the chip surface, which record different cell reactions. Five ion-sensitive field-effect transistors (ISFET) measure the pH value of the extracellular medium. A high acidification of the medium correlates with a high metabolic rate of the cells. One interdigitated electrode structure (IDES) is used to draw conclusions about the cell morphology and cell adhesion of the cells on the chip surface \citep{Thedinga2007, Ceriotti2007}. The signals of each sensor are recorded concurrently over time, representing the above aspects of the cells' metabolism. Since the latter is undoubtedly a continuous process, the sensor signals might be taken for functional data, observed at equidistant, discrete time points. In our study, we use the arithmetic mean of signals of the same type.\\
For a detection layer, we use chinese hamster lung fibroblast cells due to their stable and reliable growth \citep{Bohrn2013}. Their usual environment is nutrient medium. If the composition of this medium alters, for example by adding some test substance to it, the cells will react accordingly. To evaluate the performance of our approach in a binary classification problem, the cell chip data is restricted to measurements with nutrient medium only, and measurements where 2.5mM paracetamol is added. The data set includes $n=n_0+n_1=120$ measurements per signal type of $Q=89$ equidistant observation points, $n_0=63$ without and $n_1=57$ with AAP, depicted in Figure \ref{fig:cellsignals_params}.

The measurements can be divided into three phases: the first corresponds to an acclimatisation phase with medium flowing over the cells to let them adapt to the system and get stable physiological signals. At the second phase from about 220 minutes on, the test substance (AAP) reaches the cells. From about 400 minutes on, 0.2\% Triton X-100 is added to the medium in the last measurement phase, removing the cells from the chip surface. This last step is necessary to obtain a negative control. After the cells have adapted to their environment, shortly before the test substance is applied, one expects the cells to exhibit 100\% viability. Thus, all signals were standardized in such a way that, at about 170 minutes, the signals have a value of 100. \\

Since the ISFET- and the IDES-signals originate from different biochemical processes and measurement principles, it is adequate to treat them as two functional covariate types, such that the modeling implies multiple covariates. The character of the single curves, however, is similar, exhibiting all three measurement phases, which is the reason why identical semi-metrics with the same $k$-nearest-neighbor parameter tuples can be used for both signals.

We use all the semi-metrics listed in Table \ref{tab:semmets}. Further parameters are the numbers of nearest neighbors $k\in\mathcal{K}_{nN}=\{1,5,11,21\}$, and orders of derivation $a\in\{0,1,2\}$. The choices of $\mathbb{D}_{small}$, $\mathbb{D}_1$, $\mathbb{D}_2$ and $t_e$ mainly reflect curve regions where the AAP reaches the cells in phase two and the changeover of phase two and three. For the semi-metric $d_{\mathbb{D}_{small}}^{shortEucl}$, one of the intervals $[t_1, t_{35}]$, $[t_{36}, t_{40}]$, $[t_{41}, t_{64}]$, $[t_{65}, t_{69}]$, or $[t_{70}, t_{89}]$ is used for $\mathbb{D}_{small}$; for semi-metric $d_{bo}^{Jump}$, one of the sets $\{36,39\}$ or $\{65,68\}$ is used for $\{t_b,t_o\}$; for semi-metric $d^{relAreas}$, $\mathbb{D}_{1}$ is one of the intervals $[t_1, t_{35}]$ or $[t_{41}, t_{64}]$, and $\mathbb{D}_{2}=[t_{41}, t_{64}]$; for semi-metric $d^{Points}$, an equidistant grid $t_e=t_{mQ/10}$, $m=1,\ldots 10$, is used; and the function $\phi_{\tau}(t)=\frac{300} {\text{max} \left(\phi_{1,\tau}(t)\right)} \phi_{1,\tau}(t)$ with $\phi_{1,\tau}(t)= \frac{1}{\sqrt{2\pi}\sigma} \exp^{-\frac{1}{2}\left(\frac{t-\tau} {\sigma}\right)^2}$, $\sigma=10$ and $\tau\in\{3.120$, $45.120$, $87.120$, $135.120$, $177.120$, $219.129$, $267.129$, $309.129$, $351.129$, $399.140$, $441.140$, $483.140$, $531.140\}$ is used in semi-metric $d_{\tau}^{Scan}$. In Figure \ref{fig:cellsignals_params}, the weight functions $\phi_{\tau}(t)$ per $\tau$, used in $d_{\tau}^{Scan}$, are depicted as light gray, dotted lines; the impact points $t_e$ used in $d^{Points}$ are marked by black boxes.


\subsection{Results}
\label{sec:appl_cells_res}
With the above parameter choices, the ensemble comprises $p=1248$ ensemble members, i.e.~624 coefficients per signal type that have to be estimated. The respective single posterior probability estimates $w_{igl}$ per curve $x_{i}(t)$ are calculated as described in Section \ref{sec:method_fknn}. Afterwards, the standard deviation $\text{sd}(v_{igl})$ is calculated. If $\text{sd}(v_{igl}) \equiv 0$ $\forall i$ for a certain tuple $l$, the respective tuple is removed from the data set, since it does not contain any information concerning the class. In that way, 80 tuples are removed.

\begin{figure}[]
\centering
\includegraphics[keepaspectratio,width=.5\textwidth]{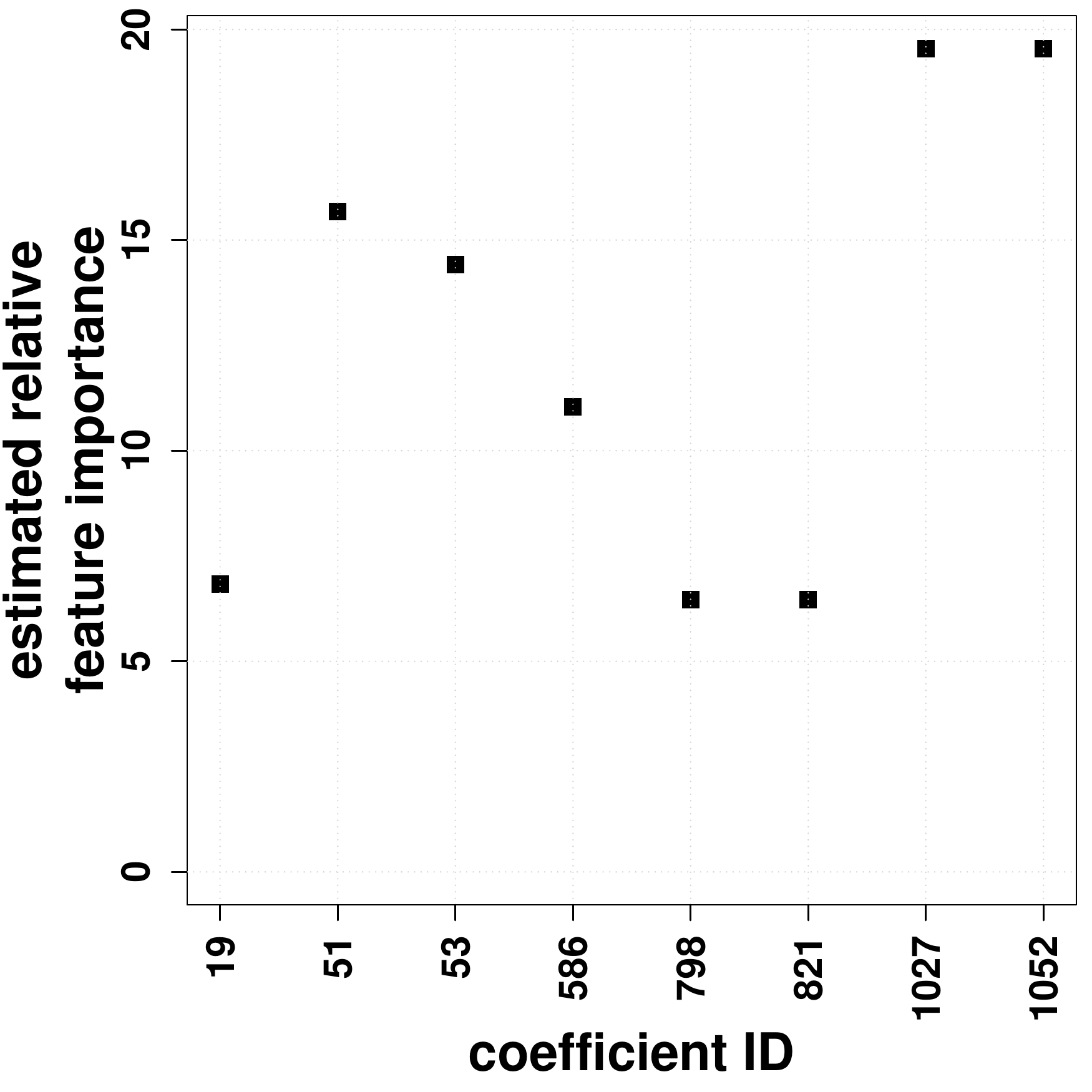}
\caption[Estimated relative feature importance of the penalized cMLM coefficients of the cell chip data.]{\small{Relative feature importance of the coefficients which have been estimated unequal to zero, as estimated from the whole cell chip data.}}
\label{fig:coeffs_cells}
\end{figure}

Finally, the penalized cMLM is applied to the whole cell chip data as described in Section \ref{sec:method_MLM}, using a global Lasso penalty and yielding a final number of $1168$ coefficients that have to be estimated. The model yielding minimal mean AIC employs a penalty parameter of $\lambda_{Lasso}$ $=$ 1.9. The RFI of the coefficient estimates resulting from the whole cell chip data set are depicted in Figure \ref{fig:coeffs_cells}. For clarity, only the $8$ coefficients that have been estimated with values unequal to zero are shown. The tuples corresponding to the five coefficients estimated with the highest RFI~values are decoded in Table \ref{tab:coeff_decode_cells}.\\

\begin{table}[]
\footnotesize
\begin{center}
	\begin{tabular}{ll}
	\toprule
	  coefficient ID & \\
	  (estimated RFI~values) & parameter tuple\\
		\midrule
		\midrule
	  Lasso & \\
		\midrule
		\begin{tabular}{l}
		1027 (19.55\%)\\
			\\
		1052 (19.55\%)\\
			\\
		51 (15.68\%)\\
		53 (14.41\%)\\
			\\
		586 (11.03\%)\\
			\\
		\end{tabular} & 
		\begin{tabular}{l}
			\{$d^{shortEucl}$, $a=2$, $k=5$\}, $\mathbb{D}_{small}=[t_{36}, t_{40}]$,\\
				covariate IDES\\
			\{$d^{shortEucl}$, with $x_i(t)$ centered, $a=2$, $k=5$\}, \\
			$\mathbb{D}_{small}=[t_{36}, t_{40}]$, covariate IDES\\
			\{$d^{Eucl}$, $a=0$, $k=5$\}, covariate ISFET\\
			\{$d^{shortEucl}$, $a=0$, $k=5$\}, $\mathbb{D}_{small}=[t_{36}, t_{40}]$,\\
			covariate ISFET\\
			\{$d^{shortEucl}$, $a=0$, $k=1$\}, $\mathbb{D}_{small}=[t_{1}, t_{35}]$,\\
			covariate IDES\\
		\end{tabular} \\
	\bottomrule
	\end{tabular}
\end{center}
\caption[Decoding of the estimated penalized cMLM coefficients of the cell chip data.]{\small{First column: the IDs of the five estimated coefficients that show the largest relative feature importance (RFI) values (in brackets, in decreasing order). Second column: The chosen ensemble coefficients are decoded, with value $a$ indicating the order of derivation and $k$ indicating the number of nearest neighbors used.}}
\label{tab:coeff_decode_cells}
\end{table}

To evaluate the prediction performance of our method and to be able to compare it to the other classification methods, we divide the data set randomly 100 times into learning sets comprising 90 curves and test sets of size 30. All competing methods are listed in Table \ref{tab:meth_tab}, including their abbreviations. They are applied on identical sample sets. 

\begin{table}
\small
\centering
	\begin{tabular}{p{4.5cm}p{4cm}p{4cm}}
	\toprule
	Method & Abbreviation & \texttt{R} function used\\
	 & & (package-name)\\
	\midrule
	Constrained multinomial & cMLM & see \cite{MRSP2015}\\
	logit model &  & and the online\\
	(global Lasso penalty) &  & supplement\\
	Constrained multinomial & cs cMLM & see above\\
	logit model (category- &  & \\
	specific Lasso penalty) &  & \\
	Constrained multinomial & csCATS cMLM & see above\\
	logit model (category- &  & \\
	specific CATS penalty) &  & \\
	Functional $k$-nearest- & kNN Ensemble & see online supplement in \\
	neighbor ensemble &  & \cite{Fuchs2015}\\
	Nonparametric functional & NPFC-deriv & funopadi.knn.lcv \\
	classification (NPFC) & & (http://www.math.univ-toulouse.fr/staph/npfda/)\\
	NPFC & NPFC-Fourier & see above \\
  NPFC & NPFC-mplsr & see above \\
	NPFC & NPFC-pca & see above \\
	Functional linear model & FLM-log & gam (mgcv)\\
	Functional random forests & fRF & FuncRandomForest (provided by the authors of \cite{Moeller2016}) \\
	Linear discriminant & LDA & lda (MASS) \\
	analysis &  &  \\
	Penalized discriminant & PDA-cov. & fda (mda) \\
	analysis &  &  \\
	Multinomial model & mM & maxent (maxent) \\
	Support vector classifiers & SVM-FPCs & svm (e1071) \\
	Support vector classifiers & SVM-cov. & see above \\
	Random forests & RF-cov. & randomForest \\
	 & & (randomForest) \\
	Random forests & RF-FPCs & see above \\
	Regularized discriminant & RDA & rda (rda) \\
	analysis &  &  \\
	Shrinkage discriminant & SDA & sda (sda) \\
	analysis &  &  \\
	\bottomrule
	\end{tabular}
\caption[List of all methods inlcuded in the comparison.]{\small{List of all methods included in the comparison.}}
\label{tab:meth_tab}
\end{table}

As with the whole data set, the estimation results of our approach per draw are sparse. Overall, only 84 coefficient IDs across all replications were estimated to have values above zero, and most were selected very seldom, see also the exemplifying boxplots in Figure \ref{fig:coeffs_cells_allsplits} in \ref{sec:app_coeffs_cells}. The coefficient ID's with the five highest estimated mean RFI~values are 1027, 1052, 586, 597, and 1043, partly overlapping with the coefficients that have been estimated from the whole data listed in Table \ref{tab:coeff_decode_cells}. The tuples corresponding to the unlisted IDs are $597$ $\hat{=}$ $\{d^{Points}$, $a=0$, $k=1$\} with covariate IDES and $1043$ $\hat{=}$ $\{d^{Scan}$, $a=2$, $k=5$\} with $\tau=219.13$ and covariate IDES. From the selected coefficients, one can conclude that both signal types, ISFET- as well as IDES-signals, imply discriminative information. Thus, the classification task is fulfilled without variable selection. Concerning feature selection, most of the selected coefficient IDs, i.e.~parameter tuples, include the curve region around 220 minutes. This is reasonable since, at this time, the AAP reaches the cells. Coefficient 586, representing a tuple including the first measurement phase of the IDES-signals, seems also a sensible choice, since many of the IDES curves show a class-dependent slope. In conclusion, the feature selection of our penalized cMLM agrees very well with background knowledge of the cell chip data.\\

\begin{figure}[]
\centering
\includegraphics[keepaspectratio,width=.9\textwidth]{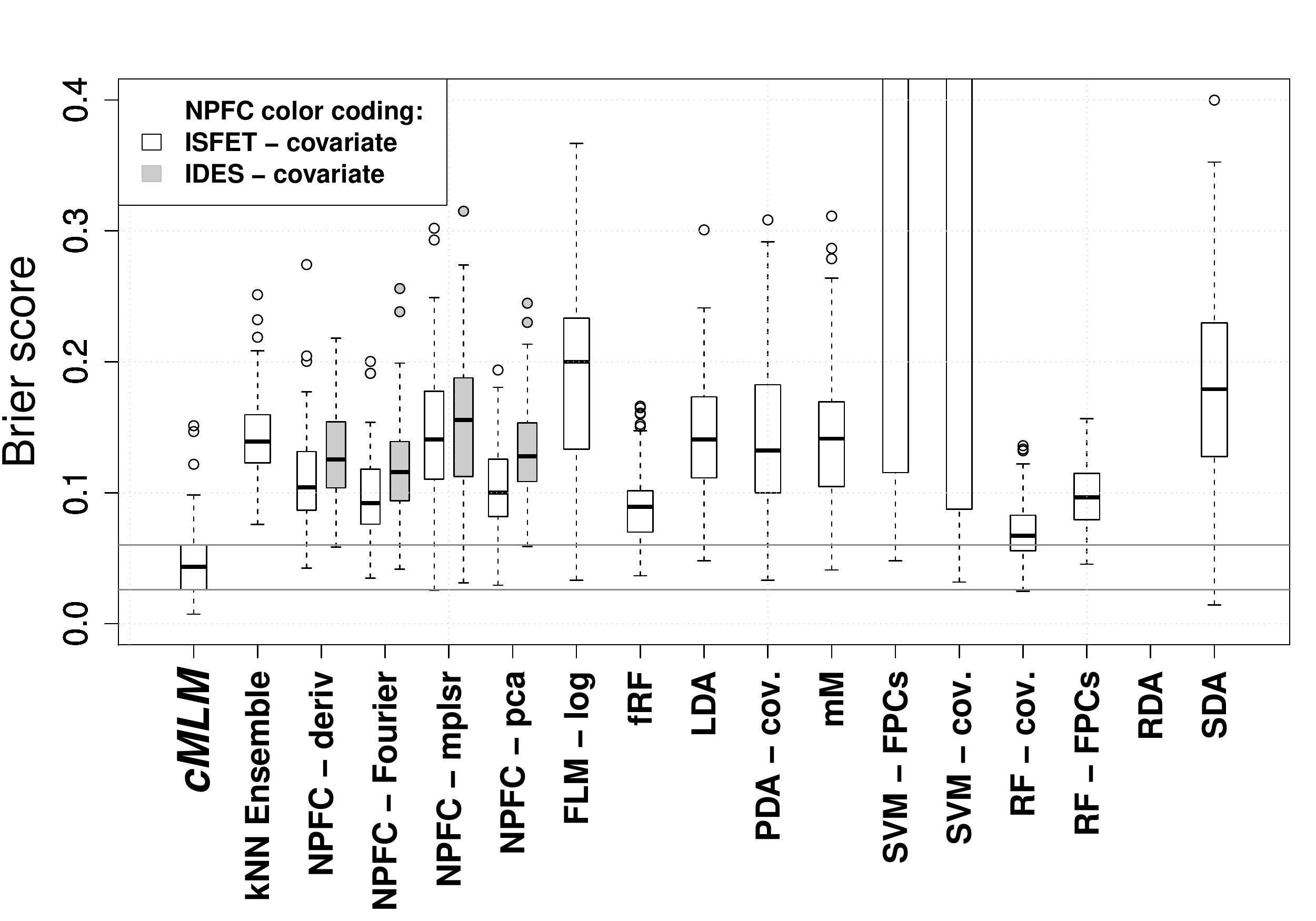}
\includegraphics[keepaspectratio,width=.9\textwidth]{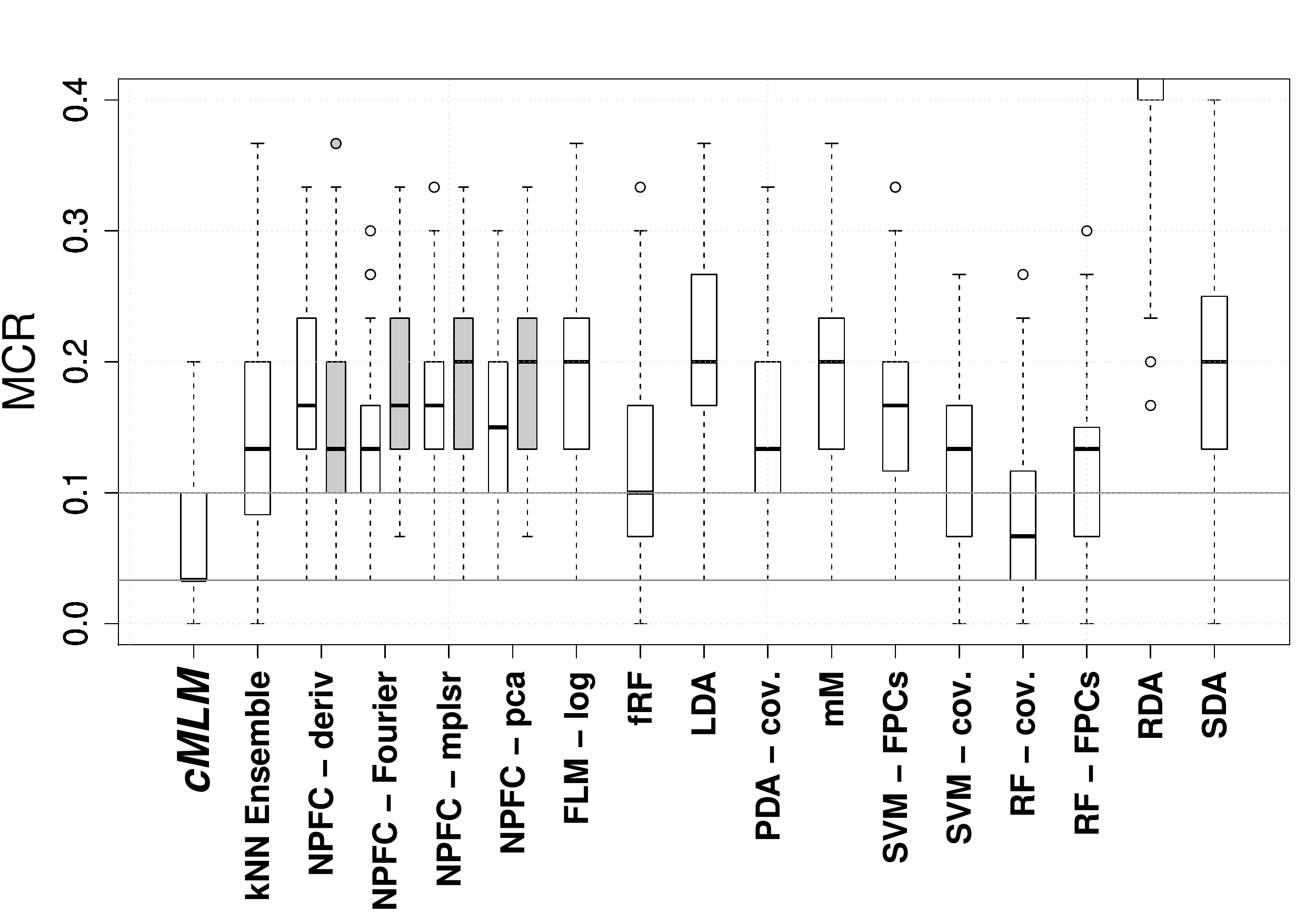}
\caption[Boxplots of the Brier scores and misclassification rates of the cell data.]{Test results of the cell data for all classification approaches on basis of 100 replications. The upper panel shows the Brier scores, the lower panel the misclassification rates (MCR). The horizontal lines indicate the values of the first and third quartiles of the cMLM-box.}
\label{fig:cells_BS_MCRs_val}
\end{figure}

Figure \ref{fig:cells_BS_MCRs_val} gives the test results for all classification approaches on basis of 100 modeling replications. The upper panel shows boxplots of the Brier score across all draws (for \textit{RDA}, the estimated probabilities were not accessible), the lower panel shows boxplots of the misclassification rates. Obviously, the penalized cMLM outperforms all other methods in terms of both performance measures. Thus, the penalized cMLM is a very attractive choice for this discrimination task.


\section{Application to real world data - phoneme data}
\label{sec:appl_phonemes}
\begin{figure}[]
\centering
\includegraphics[keepaspectratio,width=.9\textwidth]{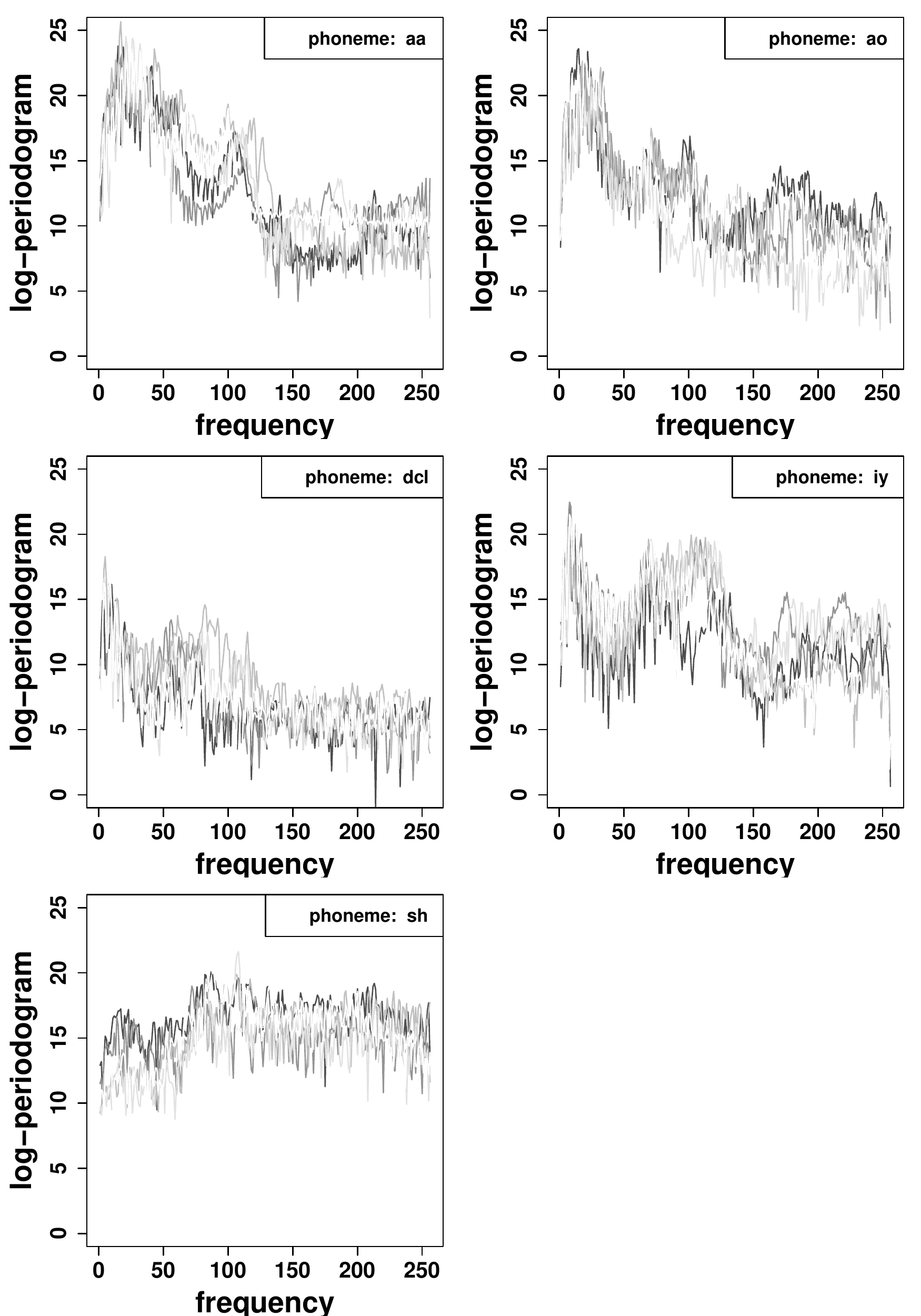}
\caption[Exemplarily curves of the phoneme data.]{\footnotesize{Five exemplarily log-periodograms per phoneme.}}
\label{fig:examples_phonemes}
\end{figure}

Another data example that became quite popular in functional data classification is the phoneme data introduced in \cite{Hastie1995}, available through the \texttt{R}-package \texttt{ElemStatLearn} \citep{elem2015}. The data consists of 4509 log-periodograms, taken as covariates $x_i(t)$, that each are ascribed to one of the five phonemes ``aa", ``ao", ``dcl", ``iy", and ``sh", recorded at 256 frequencies, i.e.~observation points. Figure \ref{fig:examples_phonemes} shows five exemplarily curves per phoneme. The goal of this discrimination task is to differ between the log-periodograms, a task arising in the field of speech recognition. The data has, among others, been examined by \cite{Hastie1995}, \cite{Ferraty2003}, \cite{Epifanio2008}, and \cite{Li2008}.\\

The parameter settings for the $k$-nearest-neighbor ensemble members were chosen arbitrarily due to the absence of relevant background knowledge. We thus use $k\in\mathcal{K}_{nN}=\{1,5,11,21\}$ nearest neighbors and orders of derivation $a\in\{0,1,2\}$. One of the intervals $[t_1, t_{17}]$, $[t_{18}, t_{36}]$, $[t_{37}, t_{56}]$, $[t_{57}, t_{76}]$, $[t_{77}, t_{100}]$, or $[t_{30}, t_{65}]$ is used for $\mathbb{D}_{small}$ in semi-metric $d_{\mathbb{D}_{small}}^{shortEucl}$; for semi-metric $d_{bo}^{Jump}$, one of the sets $\{15,19\}$, $\{34,40\}$, $\{54,58\}$, or $\{74,78\}$ is used for $\{t_b,t_o\}$; for semi-metric $d^{relAreas}$, $\mathbb{D}_{1}$ is one of the intervals $[t_1, t_{17}]$, $[t_{57}, t_{76}]$, or $[t_{30}, t_{65}]$, and $\mathbb{D}_{2}=[t_{37}, t_{56}]$; for semi-metric $d^{Points}$, a grid $t_e\in \{1$, $14.42$, $27.84$, $41.26$, $54.68$, $68.11$, $81.53$, $94.95$, $108.37$, $121.79$, $135.21$, $148.63$, $162.05$, $175.47$, $188.89$, $202.32$, $215.74$, $229.16$, $242.58$, $256\}$ is used; and the function $\phi_{\tau}(t)=$ $\frac{300} {\text{max} \left(\phi_{1,\tau}(t)\right)}$ $\phi_{1,\tau}(t)$ with $\phi_{1,\tau}(t)= \frac{1}{\sqrt{2\pi}\sigma} \exp^{-\frac{1}{2}\left(\frac{t-\tau} {\sigma}\right)^2}$, $\sigma=6$ and $\tau\in\{1$, $22$, $25$, $43$, $50$, $64$, $75$, $86$, $107$, $110$, $128$, $149$, $171$, $175$, $192$, $213$, $234$, $256\}$ is used for semi-metric $d_{\tau}^{Scan}$.


\subsection{Results}
\label{sec:appl_phonemes_res}
The upper parameter setting results in $816$ ensemble coefficients. Using the whole phoneme data, one finds 16 tuples yielding standard deviations $\text{sd}(v_{igl}) \equiv 0$ $\forall i$. These tuples are removed such that the ensemble contains $p=800$ coefficients that have to be estimated. Since the data at hand is multi-class data, category-specific penalties might be useful. Thus, we test the performance of the penalized cMLM using the global, the category-specific (cs) and the category-specific CATS (csCATS) penalties introduced in Section \ref{sec:method_MLM_penals}.\\

\begin{figure}[]
\centering
\includegraphics[keepaspectratio,width=1\textwidth]{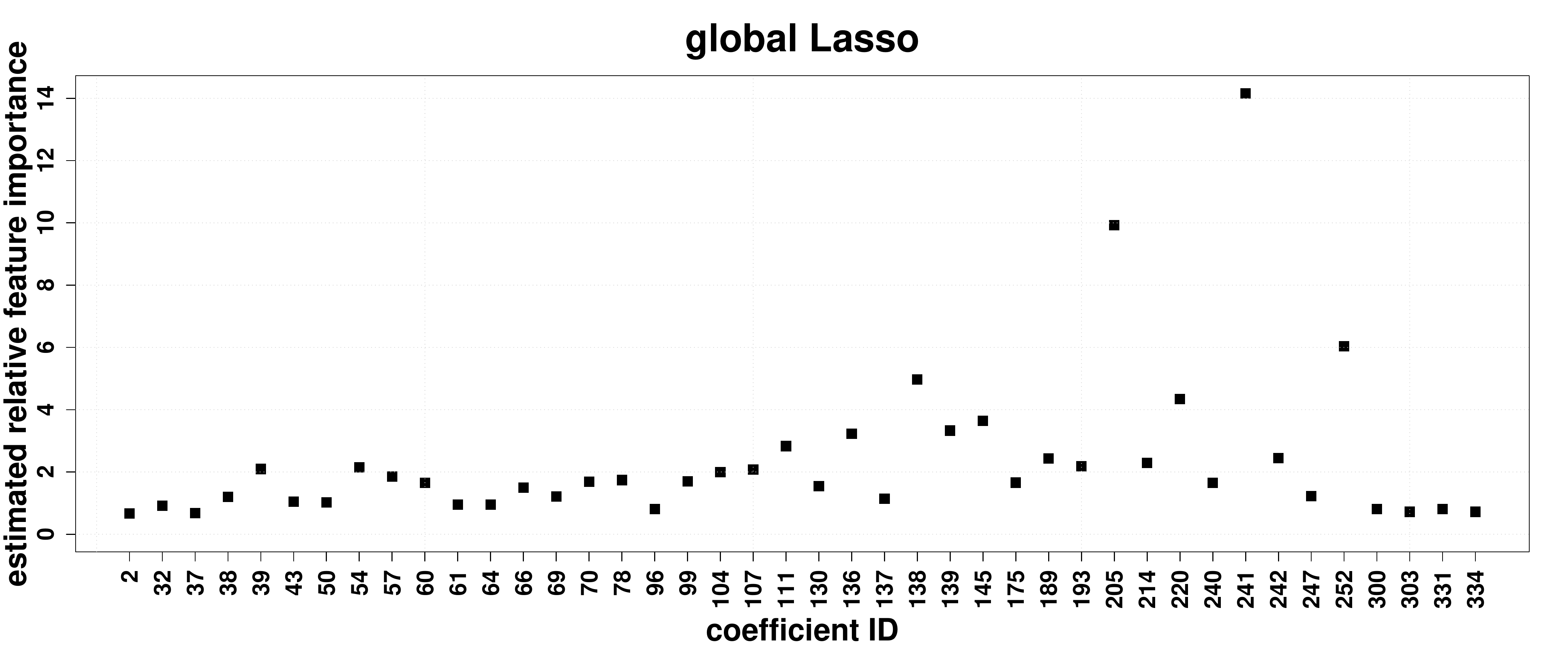}
\includegraphics[keepaspectratio,width=1\textwidth]{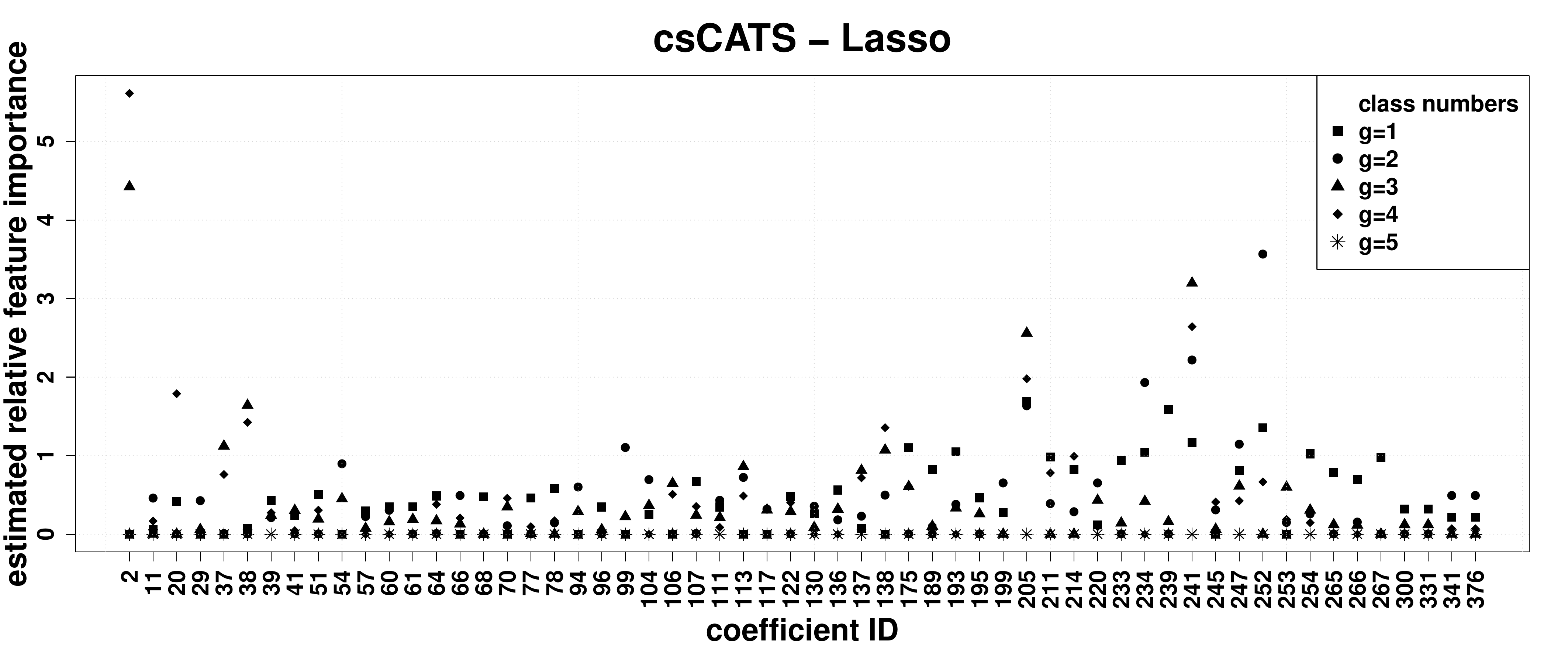}
\caption[Estimated relative feature importance of the penalized cMLM coefficients (global and cat.spec. CATS Lasso penalty) of the phoneme data.]{\small{Relative feature importance of the coefficients which have been estimated unequal to zero, as estimated from the whole phoneme data. The upper panel shows the results for the (global) Lasso penalty, the lower panel those for the csCATS penalty.}}
\label{fig:coeffs_phonemes_gl_csCATS}
\end{figure}

\begin{figure}[]
\centering
\includegraphics[keepaspectratio,width=1\textwidth]{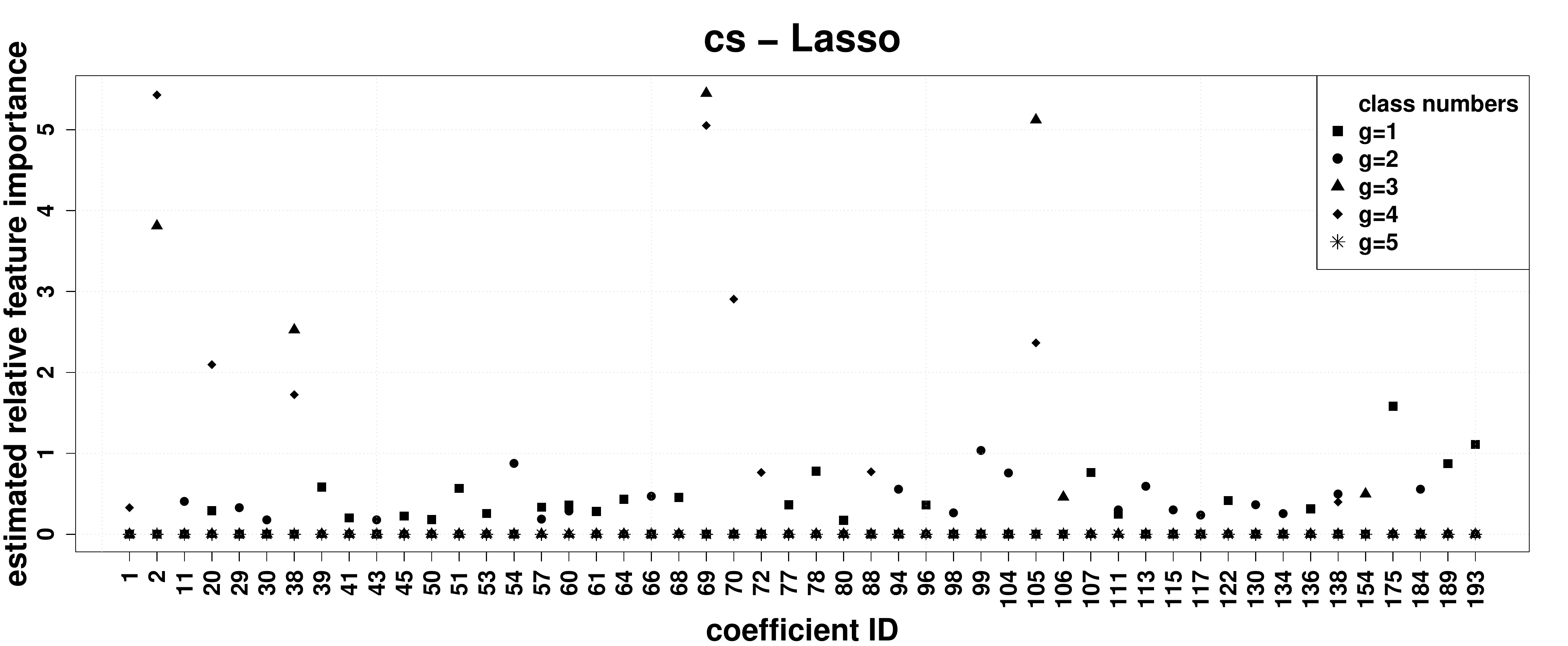}
\includegraphics[keepaspectratio,width=1\textwidth]{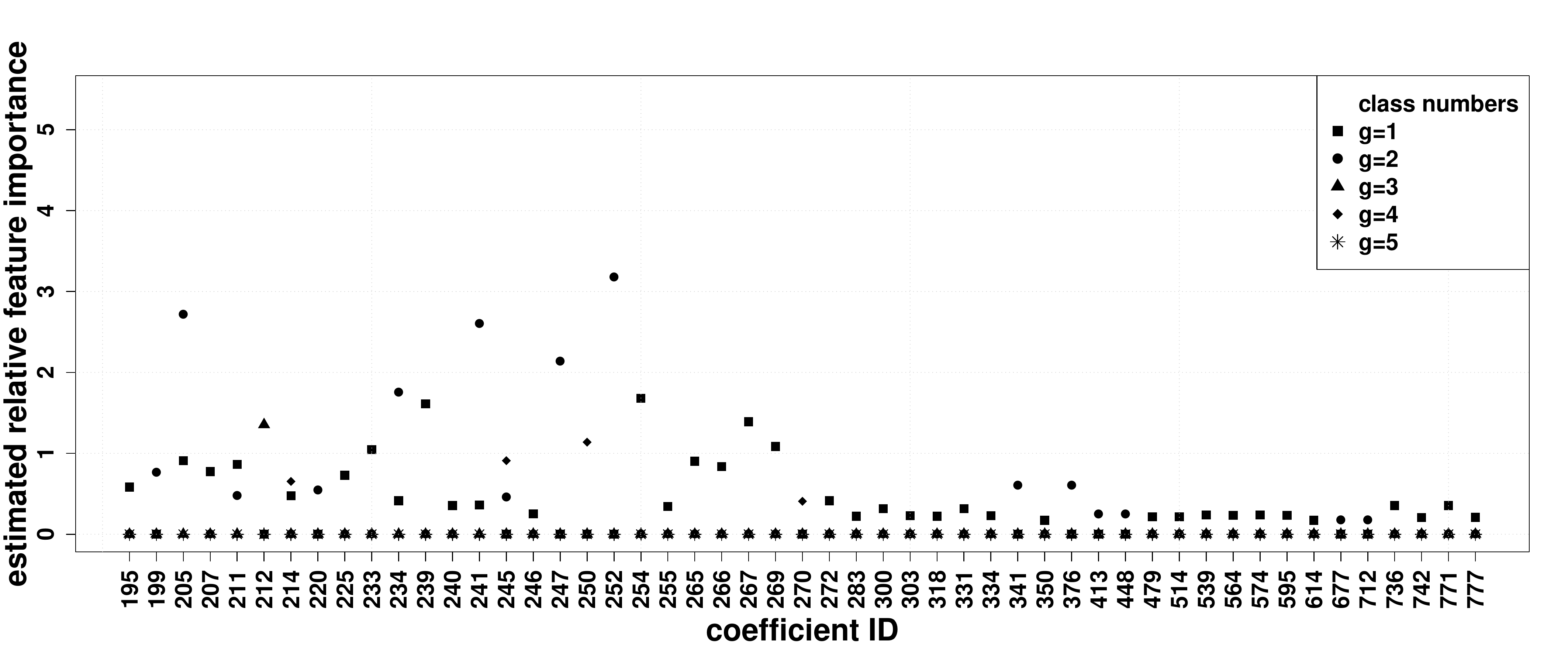}
\caption[Estimated relative feature importance of the penalized cMLM coefficients (cat.spec. Lasso penalty) of the phoneme data.]{\small{Relative feature importance of the coefficients which have been estimated unequal to zero, as estimated from the whole phoneme data, using the cs-Lasso penalty. }}
\label{fig:coeffs_phonemes_cs}
\end{figure}

In Figures \ref{fig:coeffs_phonemes_gl_csCATS} and \ref{fig:coeffs_phonemes_cs}, the symbols present the model coefficients' RFI~as estimated from the whole phoneme data, using the respective penalties. For clarity, solely the (Lasso/ cs-Lasso/ csCATS-Lasso) $\hat{=}$ (42/ 101/ 58) coefficients that are estimated to be of values unequal to zero (for at least one category in the case of cs- or csCATS-Lasso) are shown. For the two category-specific penalties, most estimated RFI~values vary strongly between the single classes. This indicates that a category-specific penalty is adequate for this data. The decoding of the four coefficients with the highest estimated RFI~values can be found in Table \ref{tab:coeff_decode_phonemes}.\\

\begin{table}[]
\footnotesize
\begin{center}
	\begin{tabular}{ll}
	\toprule
	  coefficient ID & \\
	  (estimated RFI~values) & parameter tuple\\
		\midrule
		\midrule
	  Lasso\\
		\midrule
		\begin{tabular}{l}
		241 (14.16\%)\\
		205 (9.92\%)\\
		252 (6.04\%)\\
		138 (4.97\%)\\
		\end{tabular} & 
		\begin{tabular}{l}
			\{$d^{Eucl}$ with $x_i(t)$ centered, $a=0$, $k=21$\}\\
			\{$d^{Eucl}$, $a=0$, $k=21$\}\\
			\{$d^{Max}$ with $x_i(t)$ centered, $a=0$, $k=21$\}\\
			\{$d^{shortEucl}$, $a=0$, $k=11$\}, $\mathbb{D}_{small}=[t_{1}, t_{17}]$\\
		\end{tabular} \\
	\bottomrule
	\end{tabular}
\end{center}
\begin{center}
	\begin{tabular}{ll}
	\toprule
	  coefficient ID & \\
	  (estimated RFI~values) & parameter tuple\\
		\midrule
		\midrule
	  cs-Lasso & \\
		\midrule
		\begin{tabular}{l}
		69 (10.5\%)\\
		2 (9.24\%)\\
		105 (7.49\%)\\
		38 (4.25\%)\\
			\\
		\end{tabular} & 
		\begin{tabular}{l}
			\{$d^{Eucl}$, $a=0$, $k=5$\}\\
			\{$d^{shortEucl}$, $a=0$, $k=1$\}, $\mathbb{D}_{small}=[t_{1}, t_{17}]$\\
			\{$d^{Eucl}$ with $x_i(t)$ centered, $a=0$, $k=5$\}\\
			\{$d^{shortEucl}$ with $x_i(t)$ centered, $a=0$, $k=1$\}, \\
				$\mathbb{D}_{small}=[t_{1}, t_{17}]$\\
		\end{tabular} \\
	\bottomrule
	\end{tabular}
\end{center}
\begin{center}
	\begin{tabular}{ll}
	\toprule
	  coefficient ID & \\
	  (estimated RFI~values) & parameter tuple\\
		\midrule
		\midrule
	  csCATS-Lasso & \\
		\midrule
		\begin{tabular}{l}
		2 (10.04\%)\\
		241 (9.22\%)\\
		205 (7.87\%)\\
		252 (5.58\%)\\
		\end{tabular} & 
		\begin{tabular}{l}
			\{$d^{shortEucl}$, $a=0$, $k=1$\}, $\mathbb{D}_{small}=[t_{1}, t_{17}]$\\
			\{$d^{Eucl}$ with $x_i(t)$ centered, $a=0$, $k=21$\}\\
			\{$d^{Eucl}$, $a=0$, $k=21$\}\\
			\{$d^{Max}$ with $x_i(t)$ centered, $a=0$, $k=21$\}\\
		\end{tabular} \\
	\bottomrule
	\end{tabular}
\end{center}
\caption[Decoding of the estimated penalized cMLM coefficients of the phoneme data.]{\small{Selection results for the single penalties. First column: the IDs of the four estimated coefficients that show the largest RFI~values, across all categories where appropriate (in brackets, in decreasing order). Second column: The chosen ensemble coefficients are decoded, with value $a$ indicating the order of derivation and $k$ indicating the number of nearest neighbors used. }}
\label{tab:coeff_decode_phonemes}
\end{table}

To again evaluate the prediction performance of our method and compare it to the other classification methods, 150 curves per class are drawn randomly from the complete data set, being used as a learning data set. The test sample contains another 250 randomly drawn curves per class. The draws, modeling and test steps were repeated 100 times, with all competing methods being applied on identical sample sets.\\
Again, the estimation results of our approach per draw and penalty type are sparse. Across the 100 replications, the penalized cMLM estimates overall 487 (Lasso)/ 689 (cs-Lasso)/ 315 (csCATS-Lasso) coefficient IDs to have values above zero. However, most coefficients are chosen seldomly, see also the examples in Figures B.10 - D.12 in \ref{sec:app_coeffs_phonemes} - \ref{sec:app_coeffs_phonemes_catspec_CATS}. The coefficient ID's with the four highest estimated mean RFI~values are 205, 206, 214, and 241 (Lasso)/ 1, 2, 69, and 70 (cs-Lasso)/ 2, 70, 205, and 241 (csCATS-Lasso), partly overlapping with the coefficients that have been estimated from the whole data shown in Table \ref{tab:coeff_decode_phonemes}. The tuples corresponding to the unlisted IDs are $1$ $\hat{=}$ $\{d^{Eucl}$, $a=0$, $k=1\}$, $70$ $\hat{=}$ $\{d^{shortEucl}$, $a=0$, $k=5$\} with $\mathbb{D}_{small}=[t_{1}, t_{17}]$, $206$ $\hat{=}$ $\{d^{shortEucl}$, $a=0$, $k=21$\} with $\mathbb{D}_{small}=[t_{1}, t_{17}]$, and $214$ $\hat{=}$ $\{d^{relAreas}$, $a=0$, $k=21$\} with $\mathbb{D}_{1}=[t_{57}, t_{76}]$. It thus seems that the Euclidian distance between the raw or centered curves, eventually using only the very first part of the signal, and with a relatively high number of nearest neighbors $k\geq 11$, is the most important curve characteristic for this discrimination task.

\begin{figure}[]
\centering
\includegraphics[keepaspectratio,width=.85\textwidth]{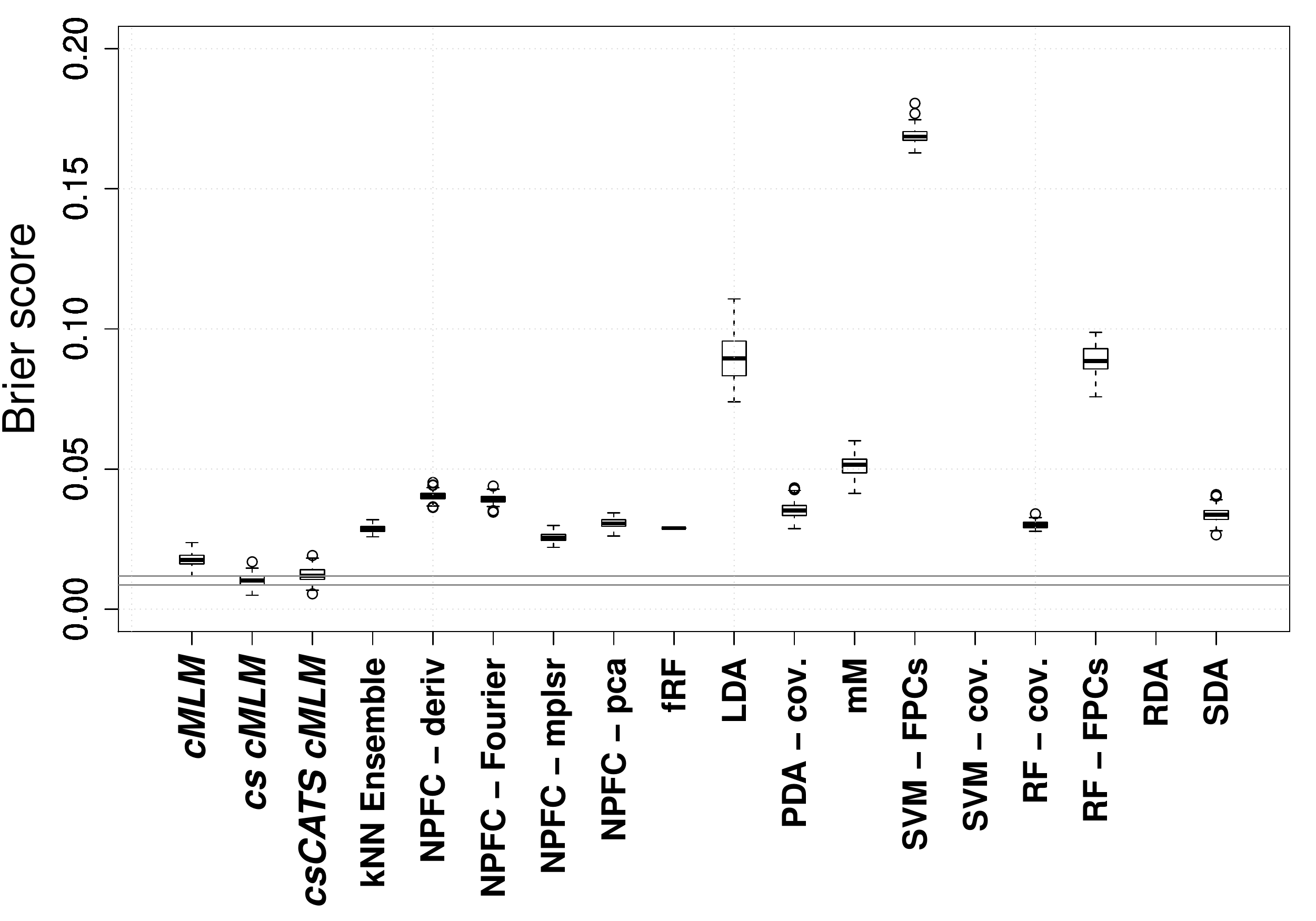}
\includegraphics[keepaspectratio,width=.85\textwidth]{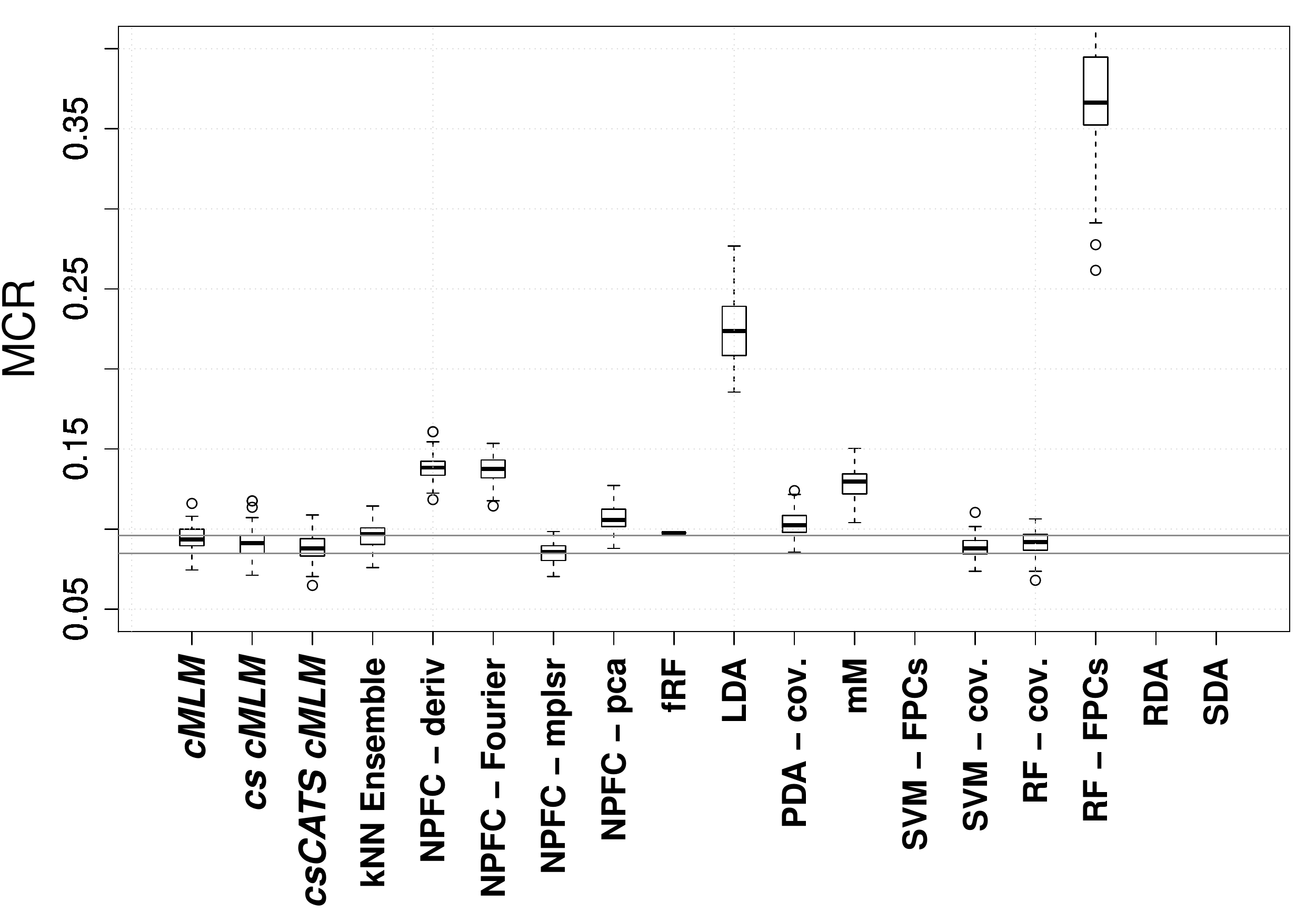}
\caption[Boxplots of the Brier scores and misclassification rates of the phoneme data.]{\small{Test results of the phoneme data for all classification approaches on basis of 100 replications. The upper panel shows the Brier scores. The box of method SVM - cov. (mean value 0.291) is not shown due to y-axis pruning. In the lower panel the misclassification rates (MCR) are shown. The boxes of SVM - FPCs, RDA and SDA are not shown due to y-axis pruning. They yield mean values of 0.8, 0.81 and 0.95, respectively. The horizontal lines indicate the values of the first and third quartiles of the category-specific cMLM-box.}}
\label{fig:phonemes_BS_MCRs_val}
\end{figure}

Figure \ref{fig:phonemes_BS_MCRs_val} gives the test results for all classification approaches on basis of 100 replications. The upper panel shows the Brier scores (for \textit{RDA}, the estimated probabilities were not accessible), the lower panel the misclassification rates. To achieve a better resolution concerning the best performing methods' boxes, the y-scale has been cut. The mean Brier score of the method SVM - cov. is 0.291. The mean MCRs of SVM - FPCs, RDA, and SDA are 0.8, 0.81, and 0.95, respectively. As can be seen, the penalized cMLM is competitive compared to the other methods for all penalties, with the lowest Brier scores and MCR values among the best methods. The Brier score also reveals an improvement of prediction accuracy compared to the $k$-nearest-neighbor ensemble, emphasizing the advantages discussed in Section \ref{sec:method}. The performance between the three penalty options Lasso, cs-Lasso and csCATS-Lasso is comparable, except for the cs-Lasso yielding somewhat lower Brier scores than the other penalties.\\


\section{Discussion}
\label{sec:discuss}
We propose a functional classification approach that includes interpretable feature and variable selection by estimating a functional $k$-nearest-neighbor ensemble within a penalized and constrained multinomial logit model. The approach represents a synthesis of the methods introduced in \cite{Tutz2015b} and \cite{Fuchs2015}. The strong performance is only obtained by the combination of the methods.\\
Setting up the functional ensemble can be seen as a dimension reduction approach that allows to detect the relevant features for classification given functional covariates. It makes efficient use of established tools and reliable and fast software. A large variety of penalties enables the user to define additional benefits arising with the discrimination itself. For example, the choice of a Lasso-type penalty results in feature selection. Since the functional $k$-nearest-neighbor ensemble is estimated within the MLM framework the estimated probabilities $\pi_{g}$ of unknown observations $x^*(t)$ are automatically scaled to their natural domain $[0,1]$, and the constraint $c_l\geq 0$ suffices to make the ensemble coefficients interpretable. This allows the use of category-specific coefficients if necessary. Moreover, the method can readily be adapted to ordinal responses. To summarize, the ensemble step enables a MLM to classify functional covariates, providing inputs that represent a wide variety of curve features. Meanwhile, the MLM step speeds up the estimation of the ensemble weights and allows for more sophisticated coefficient specification. The prediction results of both, the cell chip as well as the phoneme data, support the advantages of the method.\\

Since the number of data sets including multiple data is growing, the capability of the penalized cMLM to weight differing covariate types differently, up to selecting some types and dispense others, is of increasing interest in the context of functional classification. This was exemplified by the cell chip data. Although both functional covariate types, i.e.~the IDES- and ISFET-signals, were selected, the chosen coefficients represented different curve characteristics. Thus, even if no variable selection takes place, one gets information concerning which features of which signal provide discriminative power, allowing for insight in the processes underlying the data.\\

We have shown that the estimation of a functional $k$-nearest-neighbor ensemble via a penalized cMLM is a powerful tool for the discrimination of functional data. Nevertheless, there is ample room for extensions. Concerning the ensemble, an especially worthwhile point to be considered is the choice of the semi-metrics. To be able to achieve data-driven feature selection, a large variety of semi-metrics should be incorporated in the ensemble. The semi-metrics and respective parameters have to be chosen by the user. A randomized choice of both, semi-metrics and semi-metric parameters where appropriate, possibly from predefined sets, seems a sensible enhancement of the $k$-nearest-neighbor ensemble. Also, multivariate and non-functional covariates could be included by suitable semi-metrics.\\
Concerning the penalized cMLM, the method can handle additional covariates corresponding to (random) batch or time-independent effects by adapting the grouped Lasso penalty. There are also penalties that account for highly correlated inputs, see \cite{Tutz2009} or \cite{Bondell2008} for exemplarily penalties applied on non-functional data. However, the combination of all issues mentioned, i.e.~ordinal responses, random and time-independent effects as well as highly correlated data, being incorporated into the penalized cMLM approach will probably rise other challenges that could be topics for future research.\\


\newpage
\appendix

\newpage
\section{Estimated coefficients of the cell chip data across replication splits}
\label{sec:app_coeffs_cells}
In the main article, an important issue was the prediction performance comparison of different methods. To this end, the cell chip data was divided randomly 100 times into learning sets comprising 90 curves and test sets of size 30. While the prediction results of all competing methods were given in Figure \ref{fig:cells_BS_MCRs_val}, the following Figure \ref{fig:coeffs_cells_allsplits} gives the cumulated RFI of those penalized cMLM coefficients that are estimated unequal to zero for at least one split, as boxplots across all 100 splits. As can be seen, 84 coefficient IDs across all replications were chosen, and most were selected very seldomly.

\begin{figure}[Bh]
\centering
\includegraphics[keepaspectratio,width=1\textwidth]{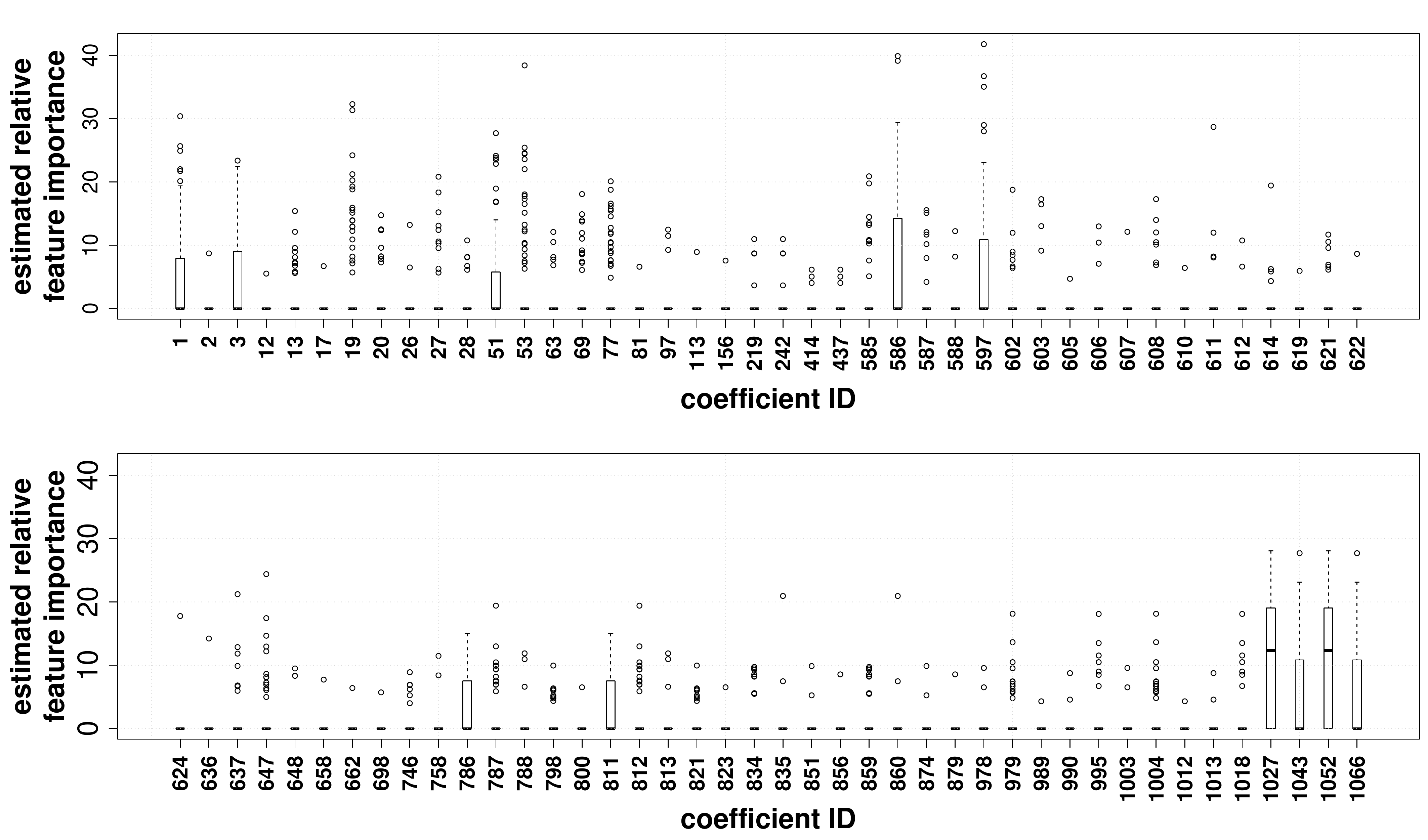}
\caption[Estimated penalized cMLM coefficients of the cell chip data across all splits.]{\small{Coefficients of the cell chip data that are unequal to zero, as boxplots across all 100 splits into learning and test data sets.}}
\label{fig:coeffs_cells_allsplits}
\end{figure}
\newpage


\section{Estimated coefficients of the phoneme data across replication splits}
\label{sec:app_coeffs_phonemes}
Analogously to the cell chip data, modeling and test steps were repeated 100 times to evaluate the prediction performance of the different classification methods. Here, each random draw used 150 curves per class as a learning data set. The test sample contained another 250 randomly drawn curves per class. The prediction results of all competing methods were given in Figure \ref{fig:phonemes_BS_MCRs_val}. The following figure gives the cumulated RFI of those globally penalized cMLM coefficients that are estimated to have mean RFI values of above 0.25\% (across all splits), as boxplots across all 100 splits. Overall, 487 from 800 coefficient IDs across all replications were chosen, most very seldomly, as, for example, the coefficients 7 - 10 exemplify.

\includepdf[pages=-,scale=.95,nup=1x1]{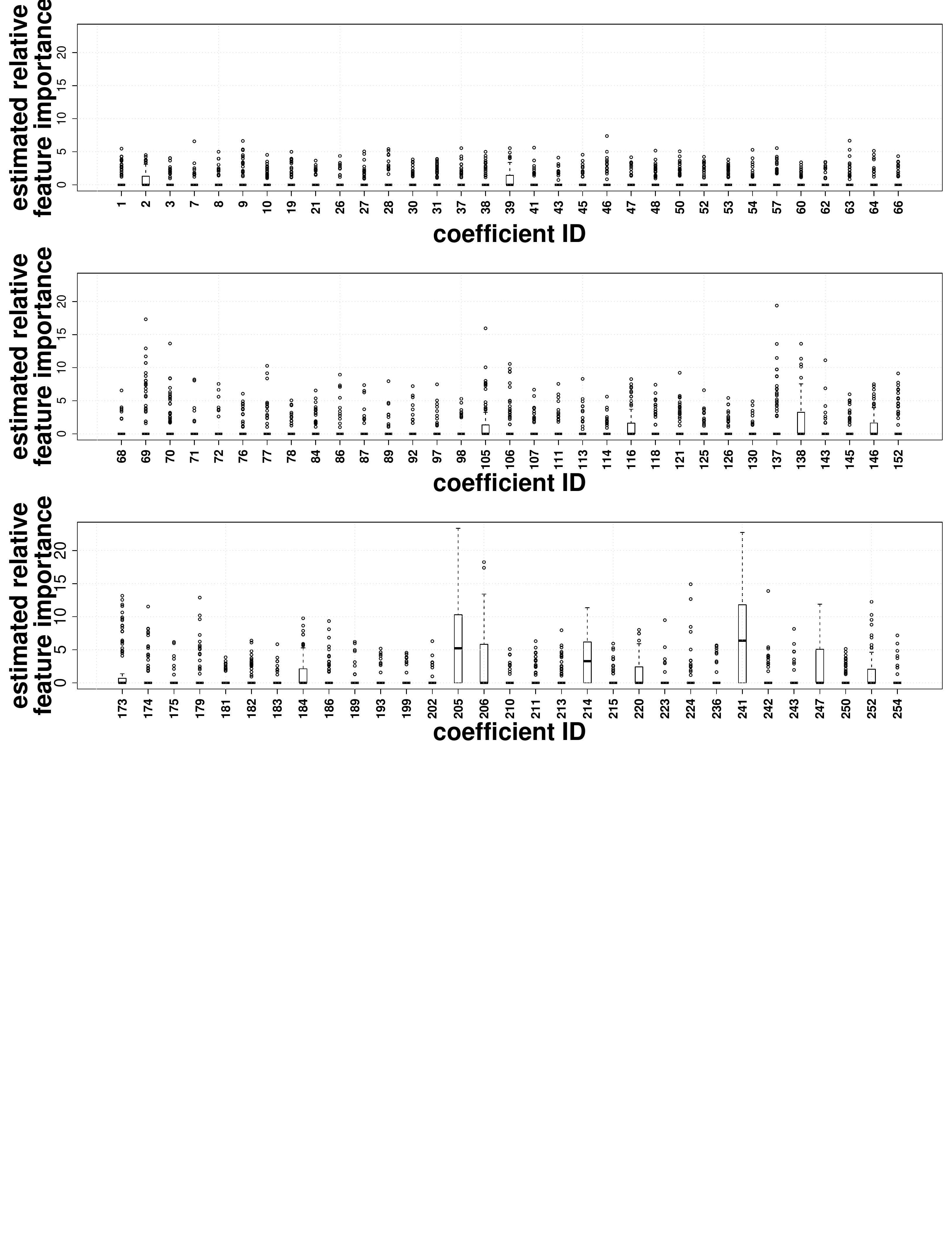}
\label{fig:coeffs_phonemes_allsplits}


\section{Estimated coefficients of the phoneme data across replication splits, using a category-specific penalty}
\label{sec:app_coeffs_phonemes_catspec}
Analogously to the cell chip data, modeling and test steps were repeated 100 times to evaluate the prediction performance of the different classification methods. Here, each random draw used 150 curves per class as a learning data set. The test sample contained another 250 randomly drawn curves per class. The prediction results of all competing methods were given in Figure \ref{fig:phonemes_BS_MCRs_val}. The following figure gives the cumulated RFI of those category-specific penalized cMLM coefficients that are estimated to have mean RFI values of above 0.25\% (across all splits, for at least one class), as boxplots across all 100 splits. Overall, 689 from 800 coefficient IDs across all replications were chosen, most very seldomly, as, for example, the coefficients 18 or 31 exemplify.
\includepdf[pages=-,scale=.95,nup=1x1]{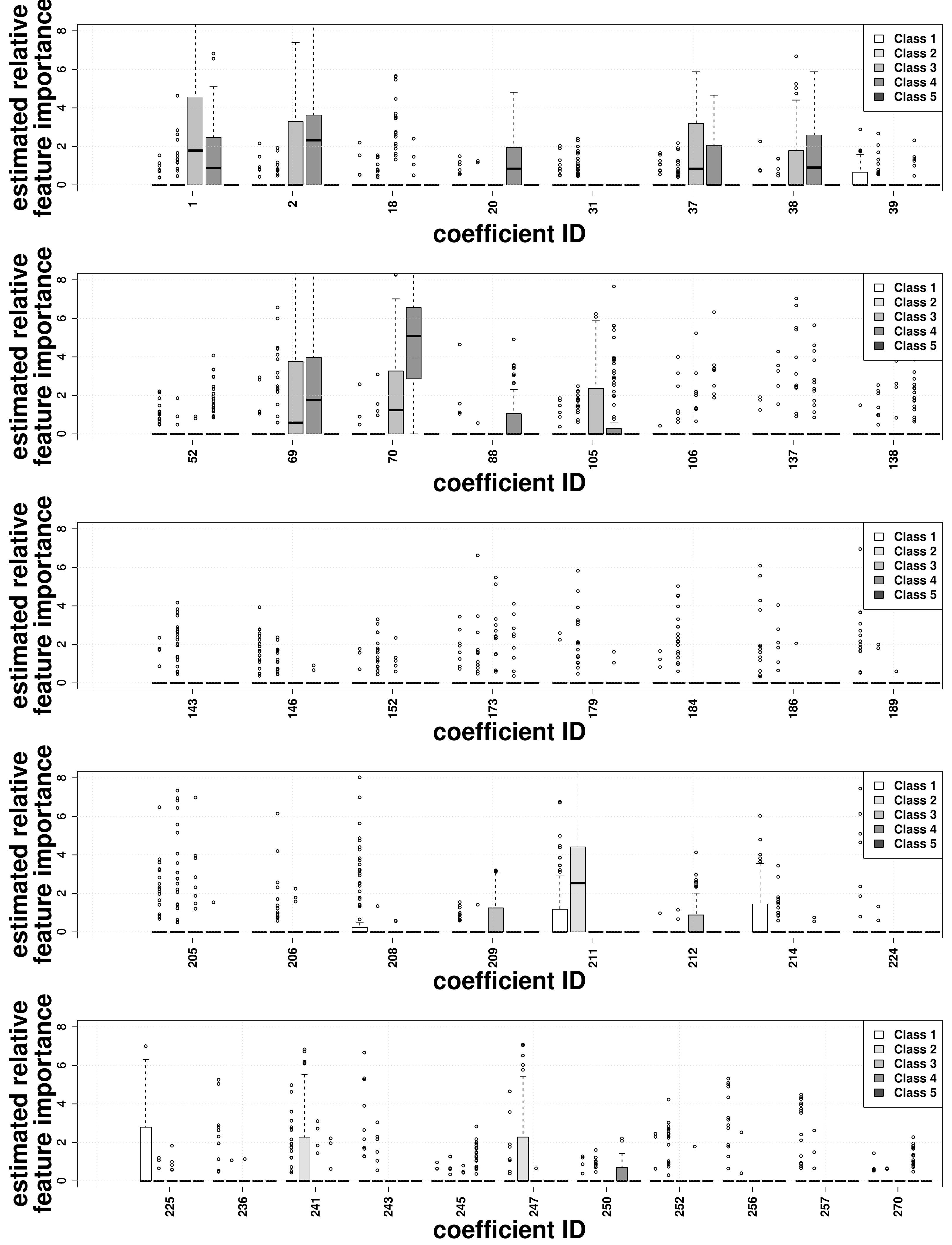}
\label{fig:coeffs_phonemes_allsplits_catspec}


\section{Estimated coefficients of the phoneme data across replication splits, using a category-specific CATS penalty}
\label{sec:app_coeffs_phonemes_catspec_CATS}
Analogously to the cell chip data, modeling and test steps were repeated 100 times to evaluate the prediction performance of the different classification methods. Here, each random draw used 150 curves per class as a learning data set. The test sample contained another 250 randomly drawn curves per class. The prediction results of all competing methods were given in Figure \ref{fig:phonemes_BS_MCRs_val}. The following figure gives the cumulated RFI of those categorically structured (CATS) penalized cMLM coefficients that are estimated to have mean RFI values of above 0.25\% (across all splits, for at least one class), as boxplots across all 100 splits. Overall, only 315 from 800 coefficient IDs across all replications were chosen, most very seldomly, as, for example, the coefficients 20 or 105 exemplify.

\includepdf[pages=-,scale=.95,nup=1x1]{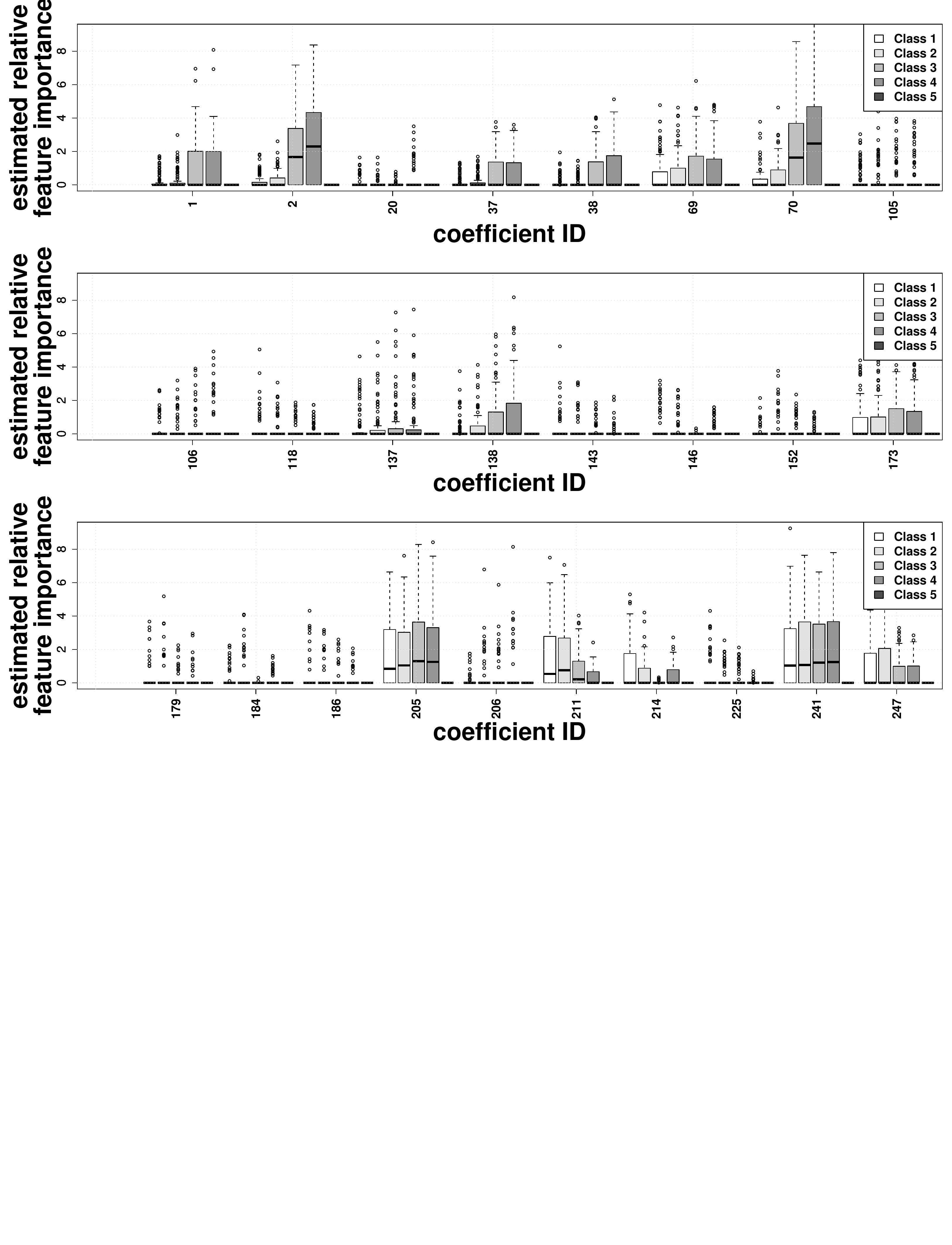}
\label{fig:coeffs_phonemes_allsplits_catspec_CATS}


\newpage
\bibliographystyle{apalike}
\bibliography{bibtest_lasso}

\end{document}